\documentclass[11pt]{article}
\usepackage{amsfonts, amsmath, amssymb, amsthm}
\usepackage[english]{babel}
\usepackage{booktabs}
\usepackage{bm}
\usepackage{eucal}
\usepackage{enumerate}
\usepackage{float}
\usepackage{dsfont} 
\usepackage{epsfig}
\usepackage{fontenc}
\usepackage[hmargin=1.6cm, vmargin=3cm]{geometry}
\usepackage{lscape}
\usepackage{mathrsfs}
\usepackage{natbib}
\usepackage{rotating}
\usepackage{verbatim}
\usepackage{xcolor}
\usepackage{dirtytalk}
\usepackage{multirow}
\usepackage{mdframed}
\usepackage[normalem]{ulem}
\usepackage{subfigure}
\usepackage{xr}
\usepackage{mleftright}
\usepackage[colorlinks,bookmarksopen,bookmarksnumbered,citecolor=blue, hyperfootnotes=false, hypertexnames=false]{hyperref}
\usepackage{titling}
\usepackage{bibunits}

\date{}

\renewcommand{\thefootnote}{$\dagger$}

\begin{document}
\begin{bibunit}[hapalike]
\title{Penalised spline estimation of covariate-specific time-dependent ROC curves}
\author{\textsc{Mar\'ia Xos\'e Rodr\'iguez-\'Alvarez} and \textsc{Vanda~In\'acio}}
\date{}
\maketitle 
\begin{abstract}
\noindent 
The identification of biomarkers with high predictive accuracy is a crucial task in medical research, as it can aid clinicians in making early decisions, thereby reducing morbidity and mortality in high-risk populations. Time-dependent receiver operating characteristic (ROC) curves are the main tool used to assess the accuracy of prognostic biomarkers for outcomes that evolve over time. Recognising the need to account for patient heterogeneity when evaluating the accuracy of a prognostic biomarker, we introduce a novel penalised-based estimator of the time-dependent ROC curve that accommodates a possible modifying effect of covariates. We consider flexible models for both the hazard function of the event time given the covariates and biomarker and for the location-scale regression model of the biomarker given covariates, enabling the accommodation of non-proportional hazards and nonlinear effects through penalised splines, thus overcoming limitations of earlier methods. The simulation study demonstrates that our approach successfully recovers the true functional form of the covariate-specific time-dependent ROC curve and the corresponding area under the curve across a variety of scenarios. Comparisons with existing methods further show that our approach performs favourably in multiple settings.  Our approach is applied to evaluate the ability of the Global Registry of Acute Coronary Events risk score to predict mortality over different time periods after discharge in patients who have suffered an acute coronary syndrome and to investigate how this ability may vary with the left ventricular ejection fraction. An \texttt{R} package, \texttt{CondTimeROC}, implementing the proposed method is provided.
\end{abstract}

\let\thefootnote\relax\footnotetext{Mar\'ia Xos\'e Rodr\'iguez-\'Alvarez, CITMaga and Departamento de Estat\'istica e Investigaci\'on Operativa, Universidade de Vigo, Vigo, Spain (\textit{mxrodriguez@uvigo.gal}). Vanda In\'acio, School of Mathematics, University of Edinburgh, Scotland, UK (\textit{vanda.inacio@ed.ac.uk})}

\noindent\textsc{key words:} Biomarker, location-scale regression model, piecewise exponential additive model, penalised splines, predictive accuracy, survival analysis

\section{\large{\textsf{Introduction}}}
Prognostic markers that accurately predict disease onset, progression, or mortality are essential in clinical practice, and it is critical to rigorously evaluate their predictive accuracy before adopting them. Prospective studies are often conducted to assess the prognostic performance of a biomarker, with the marker measured at baseline and patients followed over time to observe the occurrence of the event of interest (e.g., disease onset, recurrence, or death). In this context, the disease outcome is clearly time-dependent: subjects are initially event-free but may experience the event during the follow-up period.

The receiver operating characteristic (ROC) curve is arguably the most popular graphical tool for evaluating the discriminatory ability of a continuous biomarker for a binary disease outcome. In classical ROC analysis, within the context of a diagnostic study, a subject's disease status is treated as a fixed characteristic, with the biomarker measured concurrently with the binary disease status. To account for the time lag between marker measurement and event occurrence in prognostic studies, time-dependent extensions of sensitivity, specificity, and consequently the ROC curve, have been proposed. The seminal paper is due to \cite{heagerty00}, with \cite{heagerty05} introducing a taxonomy for time-dependent measures of accuracy (i.e., sensitivity and specificity). As detailed in Section \ref{notation_prelim}, we focus on the so-called cumulative/dynamic ROC curve, which allows to evaluate how well a marker measured at baseline can distinguish between subjects who experience the event of interest and those who do not, within a given follow-up interval of the form $[0,t]$. This has been a prolific area of research over the past two decades \citep[for a review, see][and references therein]{Blanche13a, kamarudin2017}.

It is now widely acknowledged that patient characteristics can influence the discriminatory ability of a biomarker, and that failing to account for them may lead to misleading conclusions about the biomarker's performance. Understanding how covariates affect a biomarker's performance is therefore vital for identifying the subpopulations in which the biomarker can be reliably used. Although extensive literature exists on adjusting for covariates in the ROC framework for diagnostic studies, spanning parametric, semiparametric, and nonparametric methods \citep[see][for a review]{Pardo14, inacio2021}, comparatively less attention has been given to covariate adjustment in time-dependent ROC curves for prognostic settings. Most existing approaches fall within the semiparametric framework and primarily rely on proportional hazards regression models \citep[see, e.g.,][]{xiao08, Zheng2010, Zheng2012}. While these methods provide a structured modelling framework, they have two key limitations: (i) proportional hazards are assumed for both covariates and biomarker effects, and (ii) continuous covariates effects are modelled linearly. Both (i) and (ii) may not always be appropriate in practice. From a fully nonparametric perspective, the only available method is that of \cite{MX16}, which employs kernel smoothing with inverse probability of censoring weights. However, this approach is restricted to a single continuous covariate, limiting its applicability, and it exhibits high variance, particularly in small samples, which are not uncommon in prognostic studies. Recently, \cite{Sewak2025} developed an alternative modelling strategy, employing semiparametric marginal models for the biomarker and event time distributions, linked via a Gaussian copula to capture their joint dependence. This approach, while not relying on any form of proportional hazards model, still assumes linear effects of continuous covariates.

This work is motivated by the clinically relevant application in \cite{MX16}, which examined the impact of left ventricular ejection fraction (LVEF) on the prognostic value of the Global Registry of Acute Coronary Events (GRACE) risk score for predicting post-discharge mortality across several time horizons in patients with acute coronary syndrome. Although LVEF is a well-established predictor of mortality, it is not included in the GRACE score, leading to risk estimates that may be confounded by LVEF and underscoring the need for appropriate covariate adjustment. Discrepancies between the results of the nonparametric method of \cite{MX16} and those obtained using the semiparametric approach of \cite{xiao08} underscore the sensitivity of covariate-adjusted prognostic performance to modelling assumptions. To address this gap in the literature, we develop a flexible semiparametric approach for the covariate-specific time-dependent ROC curve, modelling the hazard function of event times given the biomarker and covariates using a piecewise exponential additive model \citep{Friedman82, Bender18}. This model accommodates both continuous and categorical variables and allows for different functional forms of the biomarker and covariates effects. Importantly, and unlike \cite{xiao08}, we incorporate (i) smooth nonlinear effects of continuous covariates and of the biomarker, and (ii) time-varying (potentially nonlinear) effects, thereby addressing non-proportional hazards. As in \cite{xiao08}, our estimators of sensitivity and specificity (and thus the ROC curve) require estimation of the biomarker distribution given the covariates, which we achieve using a location-scale model that incorporates smooth nonlinear effects in both the mean and variance functions. In all cases, smooth nonlinear effects are modelled using penalised splines \citep{Eilers1996, Wood2017}. Our approach therefore offers a compromise between the competing methods proposed by \cite{xiao08} and \cite{MX16}. Notably, the variance of the estimators produced by our method can be substantially lower than that of those proposed by \cite{MX16}. Our method is also computationally appealing and an additional contribution of our work is its implementation in the publicly available \texttt{R} package \texttt{CondTimeROC}, which provides a user-friendly and broadly applicable tool for analysing a wide range of events and prognostic biomarkers or scores. For completeness, \texttt{CondTimeROC} also includes implementations of the approaches by \cite{xiao08} and \cite{MX16}.

The remainder of this paper is organised as follows. In Section \ref{method}, we introduce notation, preliminary concepts, and our modelling approach for estimating the covariate-specific time-dependent ROC curve. The performance of our method is evaluated in Section \ref{sim_main} using simulated data. In Section \ref{app_main}, we apply our approach to assess the LVEF-specific and time-dependent accuracy of the GRACE risk score for predicting post-discharge mortality in patients with acute coronary syndrome. The paper concludes with a discussion in Section \ref{sec:discussion}.

\section{\large{\textsf{Covariate-Specific Time-Dependent ROC Curve}}}\label{method}
\subsection{\textsf{Notation and Preliminaries}}\label{notation_prelim}
Let $T$ denote the time to the event of interest (e.g., death), $Y$  a continuous biomarker, and $\boldsymbol{X}$ a $q$-dimensional covariate vector comprising both continuous and categorical variables. Both $Y$ and $\boldsymbol{X}$ are measured at baseline.
We assume that higher values of $Y$ are associated with an increased hazard of the event (i.e., shorter event times). In most cases, $T$ is not fully observed due to right censoring, and only $Z = \min\{T, C\}$ is observed, where $C$ is the censoring time. Therefore, for $n$ independent and identically distributed random vectors $\{(t_i, c_i, y_i, \boldsymbol{x}_i)\}_{i = 1}^{n}$, the observed data consist of $\{(z_i, \delta_i, y_i, \boldsymbol{x}_i)\}_{i = 1}^{n}$, where $z_i = \min\{t_i, c_i\}$ and $\delta_i = I\left(t_i \leq c_i\right)$ is the censoring indicator, with $I\left(\cdot \right)$ denoting the indicator function.

Time-dependent ROC curves are defined analogously to time-independent ROC curves by extending the definitions of \textit{Sensitivity} and \textit{Specificity} to incorporate the time dimension. Several time-dependent extensions have been proposed in the literature \citep{Pepe08}, with the primary difference among them being how  cases (`diseased' individuals) and  controls (`healthy' individuals) are defined. We follow the proposal of \cite{heagerty05}, extended to the conditional setting, and focus on the cumulative \textit{Sensitivity} (Se) and dynamic \textit{Specificity} (Sp)
\begin{align}
\text{Se}^\mathbb{C}(\upsilon, t \mid \boldsymbol{x}) & =  \Pr(Y > \upsilon \mid T \leq t, \boldsymbol{X} = \boldsymbol{x}),\label{cd_se}\\
\text{Sp}^\mathbb{D}(\upsilon, t \mid \boldsymbol{x}) & = \Pr(Y \leq \upsilon \mid T > t, \boldsymbol{X} = \boldsymbol{x}).
\label{cd_sp}
\end{align}
Note that with cumulative Se and dynamic Sp, the focus is on evaluating the biomarker's ability to distinguish between individuals--with a covariate vector value $\boldsymbol{x}$--who will experience the event of interest before time $t$ (cases) and those who will experience it after time $t$ (controls). By conditioning on both $\boldsymbol{X}$ and $T$, different sensitivities and specificities, and thus different discriminatory capacities, can  be obtained for each value of $\boldsymbol{x}$ and each time point $t$. 
Based on Equations \eqref{cd_se} and \eqref{cd_sp}, the covariate-specific cumulative-dynamic time-dependent ROC curve is defined as the set $\{(1 - \text{Sp}^\mathbb{D}(\upsilon, t \mid \boldsymbol{x}), \text{Se}^\mathbb{C}(\upsilon, t \mid \boldsymbol{x})), \upsilon\in \mathbb{R}\}$. In practice, the covariate-specific cumulative-dynamic time-dependent ROC curve is typically expressed in the following functional form 
\begin{equation}  
\mbox{ROC}^{\mathbb{C}/\mathbb{D}}\left(p, t \mid \boldsymbol{x}\right) = \text{Se}^\mathbb{C}\left((1-\text{Sp}^\mathbb{D})^{-1}\left(p,t \mid \boldsymbol{x}\right), t \mid \boldsymbol{x}\right), \qquad p \in \left(0,1\right),
\label{ROC_curve}
\end{equation}
where $(1-\text{Sp}^\mathbb{D})^{-1}\left(p,t \mid \boldsymbol{x}\right) = \inf\{v \in \mathbb{R}: 1-\text{Sp}^\mathbb{D}(v,t\mid\mathbf{x}) \leq p \}$. The area under the ROC curve (AUC) is the most commonly used summary measure of prognostic accuracy and is defined in this context as
\begin{equation} 
\text{AUC}^{\mathbb{C}/\mathbb{D}}\left(t\mid \boldsymbol{x}\right) = \int^{1}_{0}{\mbox{ROC}^{\mathbb{C}/\mathbb{D}}\left(p, t \mid \boldsymbol{x}\right)\text{d}p}.
\label{AUC}
\end{equation}

\subsection{\textsf{Penalised-based Estimator}}\label{pen_est}
We now present the proposed estimators for the covariate-specific cumulative Se and dynamic Sp, as defined in Equations \eqref{cd_se} and \eqref{cd_sp}, respectively. Before proceeding, note that these equations can be expressed as follows \cite[see, e.g.,][]{xiao08}:
\begin{align}
\text{Se}^\mathbb{C}(\upsilon,t \mid \boldsymbol{x}) = & \frac{\Pr(Y > \upsilon, T \leq t\mid \boldsymbol{X} = \boldsymbol{x})}{\Pr(T \leq t\mid \boldsymbol{X} = \boldsymbol{x})} = \frac{\int_{\upsilon}^{\infty}\left(1 - S_T\left(t \mid y, \boldsymbol{x}\right)\right)\text{d}F_Y\left(y \mid \boldsymbol{x} \right)}{\int_{-\infty}^{\infty}\left(1 - S_T\left(t \mid y, \boldsymbol{x}\right)\right)\text{d}F_Y\left(y \mid \boldsymbol{x} \right)},\label{cd_se_new}\\
\text{Sp}^\mathbb{D}(\upsilon,t \mid \boldsymbol{x}) = & \frac{\Pr(Y \leq \upsilon, T > t \mid \boldsymbol{X} = \boldsymbol{x})}{\Pr(T > t\mid \boldsymbol{X} = \boldsymbol{x})} = \frac{\int_{-\infty}^{\upsilon}S_T\left(t \mid y, \boldsymbol{x}\right)\text{d}F_Y\left(y \mid \boldsymbol{x} \right)}{\int_{-\infty}^{\infty}S_T\left(t \mid y, \boldsymbol{x}\right)\text{d}F_Y\left(y \mid \boldsymbol{x} \right)},\label{cd_sp_new}
\end{align}
where
\[
S_T\left(t \mid y, \boldsymbol{x}\right) = \Pr\left(T > t \mid Y = y, \boldsymbol{X} = \boldsymbol{x}\right)\;\; \mbox{and} \;\;F_Y\left(y \mid \boldsymbol{x} \right) = \Pr\left(Y \leq y \mid \boldsymbol{X} = \boldsymbol{x}\right).
\]
Thus, to estimate (\ref{cd_se_new}) and (\ref{cd_sp_new}), we simply require estimators for $S_T\left(\cdot \mid y, \boldsymbol{x}\right)$ and $F_Y\left(\cdot \mid \boldsymbol{x} \right)$. The details are provided in the next subsections.

\subsubsection{\texorpdfstring{\textbf{\textsf{Estimation of $S_T\left(\cdot \mid y, \boldsymbol{x}\right)$}}}{Estimation of ST}}
To estimate $S_T\left(\cdot \mid y, \boldsymbol{x}\right)$ we assume a regression-type model for the conditional hazard function $\lambda\left(t \mid y, \boldsymbol{x}\right)$ and make use of the following equality
\begin{equation}
S_T\left(t \mid y, \boldsymbol{x}\right) = \exp\left(-\int_{0}^{t}\lambda\left(u \mid y, \boldsymbol x\right)\text{d}u\right).
\label{S_lambda_rel}
\end{equation}
More precisely, and in its more general specification, we consider a model of the form
\begin{align}
\lambda\left(t \mid y, \boldsymbol x\right) & = \exp\left(\eta(t \mid y, \boldsymbol{x})\right)\nonumber\\& = \exp\left(\alpha_0 + h_0\left(t\right) + h_Y\left(y\right) + \sum_{a=1}^{A}h_a\left(\boldsymbol{x}_a\right) + f_{Y,T}\left(y,t\right) + \sum_{b=1}^{B}f_{b,T}\left(\boldsymbol{x}_b, t\right) + \sum_{d=1}^{D}f_{d,Y}\left(\boldsymbol{x}_d, y\right)\right),
\label{hazard_gam}
\end{align}
where $\boldsymbol{x}_a$, $\boldsymbol{x}_b$, and $\boldsymbol{x}_d$ denote subsets of the $q$ covariates in $\boldsymbol{x}$, and $h_{\{\cdot\}}$ and  $f_{\{\cdot\}}$ represent generic functional forms corresponding to different types of effects. Our proposal is flexible in the types of covariates it can accommodate, including both categorical and continuous variables, and more importantly, in the types of functional effects it allows. Beyond the standard linear and parametric effects considered in previous proposals, our approach permits smooth nonlinear effects of continuous variables in one or more dimensions (i.e., involving one or more variables), as well as smooth varying-coefficient terms. The specification of these smooth functions relies on penalised regression splines. Since penalised splines are also a key component for estimating $F_Y\left(\cdot \mid \boldsymbol{x} \right)$, we defer the details to Section \ref{smooth_fun} below. Furthermore, note that in \eqref{hazard_gam} we explicitly include the possibility of time-varying effects of the biomarker and of (some) continuous/categorical covariates through $f_{Y,T}$ and $f_{b,T}$, respectively. By incorporating these time-varying effects, we allow our regression modelling approach to accommodate non-proportional hazards.

Estimation of model \eqref{hazard_gam} is based on the Piecewise Exponential (Additive) Model \citep[PEM/PAM,][]{Friedman82, Bender18}. Through a data augmentation strategy, the PEM/PAM framework enables (penalised) Poisson-maximum likelihood estimation of model \eqref{hazard_gam} in the presence of censored observations. We outline the high-level idea of the PEM/PAM and refer the reader to \cite{Friedman82} and \cite{Bender18} for further details. Assuming non-informative censoring and conditional independence between observations, it can be shown that the full unpenalised log-likelihood \citep[][Chapter 3]{cox1988} is
\[
\mbox{LogLik} = \sum_{i = 1}^{n}\left(\delta_i \eta(z_i\mid y_i, \boldsymbol{x}_i) - \int_{0}^{z_i}\exp(\eta(t\mid y_i, \boldsymbol{x}_i))\text{d}t\right).
\]
PEM/PAM assume that the follow-up period $(0, t_{\mbox{max}}]$, where $t_{\mbox{max}}$ denotes the maximum observed follow-up time, can be divided into successive intervals
\begin{equation}
(\kappa_0, \kappa_1], (\kappa_1, \kappa_2], \ldots, (\kappa_{S-1}, \kappa_S],\quad \kappa_0 = 0,\quad \kappa_S = t_{\mbox{max}},
\label{break_points}
\end{equation}
such that all time-varying quantities in \eqref{hazard_gam} remain constant within these intervals. Accordingly,
\[
\eta(t\mid y_i, \boldsymbol{x}_i) = \eta_s(y_i, \boldsymbol{x}_i), \quad \forall t \in (\kappa_{s-1}, \kappa_s], \quad s = 1, \ldots, S,
\]
with
\begin{equation}
\eta_s(y_i, \boldsymbol{x}_i) = \alpha_0 + h_0\left(t_s\right) + h_Y\left(y_i\right) + \sum_{a=1}^{A}h_a\left(\boldsymbol x_{ai}\right) + f_{Y,T}\left(y_i,t_s\right) + \sum_{b=1}^{B}f_{b,T}\left(\boldsymbol x_{bi}, t_s\right) + \sum_{d=1}^{D}f_{d,Y}\left(\boldsymbol x_{di}, y_i\right),
\label{add_pred_pam_des}
\end{equation} 
and the full log-likelihood can be rewritten as
\begin{align}
\mbox{LogLik} & = \sum_{i = 1}^{n}\left(\delta_i \eta(z_i\mid y_i, \boldsymbol{x}_i) - \int_{0}^{z_i}\exp(\eta(t\mid y_i, \boldsymbol{x}_i))\text{d}t\right) \nonumber\\
& = \sum_{i = 1}^{n}\left(\delta_i\eta_{k_i}(y_i, \boldsymbol{x}_i) - \sum_{s = 1}^{k_i}R_{is}\exp\left(\eta_s(y_i, \boldsymbol{x}_i)\right)\right)\nonumber\\
& = \sum_{i = 1}^{n}\left(\sum_{s = 1}^{k_i}\left(\tilde{z}_{is}\eta_s(y_i, \boldsymbol{x}_i) - R_{is}\exp\left(\eta_s(y_i, \boldsymbol{x}_i)\right)\right)\right)\label{pam_loglik_ind},
\end{align}
where $k_i \in\{1,\ldots,S\}$ is such that $z_i \in (\kappa_{k_i-1}, \kappa_{k_i}]$, $R_{is} = \min\{\kappa_s - \kappa_{s-1}, z_i - \kappa_{s-1}\}$, and
\[
\tilde{z}_{is} = \left \{
\begin{array}{ll}
1 & \mbox{if individual $i$ has an event in}\;\;(\kappa_{s-1},\kappa_s],\\
0 & \mbox{if individual $i$ survives or is censored in}\;\;(\kappa_{s-1}, \kappa_s].
\end{array}
\right .
\]
A closer look at \eqref{pam_loglik_ind} reveals that it corresponds to the log-likelihood of a log-link Poisson model for the pseudo-observations $\tilde{z}_{is}$, with additive predictors $\eta_s(y_i, \boldsymbol{x}_i)$, where $t_s$ in \eqref{add_pred_pam_des} corresponds to $\kappa_s$, and offsets $O_{is} = \log(R_{is})$. As such, estimation of model \eqref{hazard_gam} can be carried out using standard methods for Poisson Generalised (Additive) regression Models (GAM) based on the sample $\{\{\tilde{z}_{is}, t_s, y_i, \boldsymbol{x}_i, O_{is}\}_{i=1}^{n}\}_{s = 1}^{k_i}$.

Once an estimate of $\lambda\left(\cdot \mid y, \boldsymbol{x} \right)$ is obtained, say $\widehat{\lambda}\left(\cdot \mid y, \boldsymbol{x} \right)$, $S_T\left(\cdot \mid y, \boldsymbol{x}\right)$ is estimated using the relation given in \eqref{S_lambda_rel}, i.e.,
\begin{equation}
\widehat{S}_T\left(t \mid y, \boldsymbol{x}\right) = \exp\left(-\int_{0}^{t}\widehat{\lambda}\left(u \mid y, \boldsymbol x\right)\text{d}u\right),
\label{S_lambda_rel_est}
\end{equation}
where the integral is approximated using the composite Simpson's rule.
\subsubsection{\texorpdfstring{\textbf{\textsf{Estimation of $F_Y\left(\cdot \mid \boldsymbol{x} \right)$}}}{Estimation of FY}}\label{est_F_y}
To estimate $F_Y\left(\cdot \mid \boldsymbol{x} \right)$, we assume a location-scale regression model for $Y$
\begin{equation*}
Y = \mu\left(\boldsymbol{x}\right) + \sigma\left(\boldsymbol{x}\right)\varepsilon,
\end{equation*}
where $\mu(\boldsymbol{x})=\mathbb{E}(Y\mid\boldsymbol{X} = \boldsymbol{x})$ and $\sigma^2(\boldsymbol{x}) = \text{var}(Y\mid\boldsymbol{X} = \boldsymbol{x})$ are the conditional mean and variance functions, respectively. The error term $\varepsilon$ is independent of $\boldsymbol{X}$, with mean $0$, variance $1$, and cumulative distribution function denoted by $F_{\varepsilon}$. We thus have the following relationship
\begin{equation}
F_Y\left(y \mid \boldsymbol{x} \right) = F_\varepsilon\left(\frac{y - \mu\left(\boldsymbol{x}\right)}{\sigma\left(\boldsymbol{x}\right)}\right),
\label{F_y_eqn}
\end{equation}
and estimating $F_Y\left(\cdot \mid \boldsymbol{x} \right)$ involves estimating $\mu\left(\cdot\right)$, $\sigma\left(\cdot\right)$, and $F_{\varepsilon}(\cdot)$. As in the model for the conditional hazard function, flexible specifications for the mean and variance functions are considered, i.e.,
\begin{equation}
\mu(\boldsymbol{x}) = \beta_0 + \sum_{m=1}^{M}f_m\left(\boldsymbol{x}_m\right)\;\;\mbox{and}\;\;
\sigma(\boldsymbol{x}) = \exp\left(\beta_1 + \sum_{l=1}^{L}g_l\left(\boldsymbol{x}_l\right)\right),
\label{loc_scale_model_2}
\end{equation}
where, as before, $f_m$ and $g_l$ may represent linear or parametric effects as well as smooth nonlinear effects modelled through penalised splines. Estimation of the regression and variance functions in \eqref{loc_scale_model_2} proceeds via a stage-wise approach, as proposed by \cite{Rodriguez2011}. The core of this approach relies on the following result
\begin{equation*}
\mathbb{E}\left[\log\{Y - \mu(\boldsymbol{x})\}^2 \mid \boldsymbol{X} = \boldsymbol{x}\right]  = \log\left\{\sigma^{2}\left(\boldsymbol{x}\right)\right\} + \mathbb{E}\left\{\log\left(\varepsilon^{2}\right)\right\} = \beta_{1}^{\prime} + \sum_{l=1}^{L}g^{\prime}_l\left(\boldsymbol{x}_l\right),
\end{equation*}
where $\beta_{1}^{\prime}=2\beta_{1} + \mathbb{E}\left\{\log\left(\varepsilon^{2}\right)\right\}$ and $g^{\prime}_l\left(\cdot\right) = 2g_l\left(\cdot\right)$. Thus
\begin{equation*}
\sigma^2(\boldsymbol{x}) = \gamma\exp\left\{\beta_{1}^{\prime} + \sum_{l=1}^{L}g^{\prime}_l\left(\boldsymbol{x}_l\right)\right\},
\end{equation*}
where $\gamma = 1/\exp\left[\mathbb{E}\left(\log\left\{\varepsilon^{2}\right\}\right)\right]$. This leads to the following three steps for estimating $\mu(\boldsymbol{x})$ and $\sigma^2(\boldsymbol{x})$
\begin{enumerate}
\item Estimate
\begin{equation*}
\mu(\boldsymbol{x}) = \beta_0 + \sum_{m=1}^{M}f_m\left(\boldsymbol{x}_m\right),
\end{equation*}
based on $\{(\boldsymbol{x}_i, y_i)\}_{i=1}^{n}$ using penalised-spline-based methods.
\item Similarly, estimate
\begin{equation*}
\mathbb{E}\left[\log\{Y - \mu(\boldsymbol{x})\}^2 \mid \boldsymbol{X} = \boldsymbol{x}\right] = \beta_{1}^{\prime} + \sum_{l=1}^{L}g^{\prime}_l\left(\boldsymbol{x}_l\right),
\end{equation*}
based on $\{(\boldsymbol{x}_i, \log[y_i - \widehat{\mu}(\boldsymbol{x}_i)]^2)\}_{i=1}^{n}$.
\item Estimate $\sigma^2(\boldsymbol{x})$ as
\begin{equation*}
\widehat{\sigma}^2(\boldsymbol{x}) = \widehat{\gamma}\exp\left\{\widehat{\beta}_1^{\prime} + \sum_{l=1}^{L} \widehat{g}_l^{\prime}(\boldsymbol{x}_l)\right\},
\end{equation*}
where $\widehat{\gamma}$ is the least squares estimator of $\gamma$, i.e.,
\begin{equation*}
\widehat{\gamma} = \frac{\sum_{i=1}^{n} \left\{(y_i -\widehat{\mu}(\boldsymbol{x}_i))^2 \exp\left(\widehat{\beta}^{\prime}_1 + \sum_{l=1}^{L}\widehat{g}^{\prime}_{l}(\boldsymbol{x}_{l,i})\right)\right\}}{\sum_{i=1}^{n}\left\{\exp\left(\widehat{\beta}^{\prime}_1 + \sum_{l=1}^{L}\widehat{g}^{\prime}_{l}(\boldsymbol{x}_{l,i})\right)\right\}^2}).
\end{equation*}
\end{enumerate}
Note that, using this procedure, estimating $\mu\left(\cdot\right)$ and $\sigma^2\left(\cdot\right)$ amounts to fitting two penalised-spline-based Additive Models (AM). After obtaining estimates of these quantities, $F_{\varepsilon}(\cdot)$ is estimated via the empirical distribution function of the standardised residuals, i.e.,
\begin{equation*}
\widehat{F}_{\varepsilon}(\epsilon) = \frac{1}{n}\sum_{i = 1}^{n}I\left(\widehat{\varepsilon}_i \leq \epsilon\right) = \frac{1}{n}\sum_{i = 1}^{n}I\left(\frac{y_i - \widehat{\mu}(\boldsymbol{x}_i)}{\widehat{\sigma}(\boldsymbol{x}_i)} \leq \epsilon\right) , \quad \widehat{\varepsilon}_i = \frac{y_i - \widehat{\mu}(\boldsymbol{x}_i)}{\widehat{\sigma}(\boldsymbol{x}_i)}.
\end{equation*}
Therefore, using Equation \eqref{F_y_eqn}, an estimate of $F_Y\left(\cdot \mid \boldsymbol{x} \right)$ is obtained as follows
\begin{equation}
\widehat{F}_Y\left(y \mid \boldsymbol{x}\right)=
\widehat{F}_{\varepsilon}\left(\frac{y - \widehat{\mu}(\boldsymbol{x})}{\widehat{\sigma}(\boldsymbol{x})} \right) = \frac{1}{n}\sum_{i=1}^{n}I\left(\widehat{\sigma}(\boldsymbol{x})\widehat{\varepsilon}_i + \widehat{\mu}(\boldsymbol{x}) \leq y\right).
\label{est_F_Y}
\end{equation}
\subsubsection{\textbf{\textsf{Estimation of the Sensitivity, Specificity, ROC Curve and AUC}}}
With all the components introduced above, estimates of the covariate-specific cumulative sensitivity  and dynamic specificity, as defined in Equations \eqref{cd_se_new} and \eqref{cd_sp_new}, respectively,  are obtained by plugging the estimates of $S_T$ and $F_Y$, given in \eqref{S_lambda_rel_est} and \eqref{est_F_Y} respectively, into these expressions. Specifically,
\begin{align}
\widehat{\text{Se}}^\mathbb{C}(\upsilon,t \mid \boldsymbol{x}) = & \frac{\int_{\upsilon}^{\infty}\left(1 - \widehat{S}_T\left(t \mid y, \boldsymbol{x}\right)\right)\text{d}\widehat{F}_Y\left(y \mid \boldsymbol{x} \right)}{\int_{-\infty}^{\infty}\left(1 - \widehat{S}_T\left(t \mid y, \boldsymbol{x}\right)\right)\text{d}\widehat{F}_Y\left(y \mid \boldsymbol{x} \right)} = \frac{\int_{\upsilon}^{\infty}\left(1 - \widehat{S}_T\left(t \mid y, \boldsymbol{x}\right)\right)\text{d}\widehat{F}_{\varepsilon}\left(\frac{y - \widehat{\mu}(\boldsymbol{x})}{\widehat{\sigma}(\boldsymbol{x})}\right)}{\int_{-\infty}^{\infty}\left(1 - \widehat{S}_T\left(t \mid y, \boldsymbol{x}\right)\right)\text{d}\widehat{F}_{\varepsilon}\left(\frac{y - \widehat{\mu}(\boldsymbol{x})}{\widehat{\sigma}(\boldsymbol{x})}\right)}\nonumber\\
= & \frac{\sum_{i=1}^{n}\left(1 - \widehat{S}_T\left(t \mid \widehat{\sigma}(\boldsymbol{x})\widehat{\varepsilon}_i + \widehat{\mu}(\boldsymbol{x}), \boldsymbol{x}\right)\right)I\left(\widehat{\sigma}(\boldsymbol{x})\widehat{\varepsilon}_i + \widehat{\mu}(\boldsymbol{x}) > \upsilon \right)}{\sum_{i=1}^{n}\left(1 - \widehat{S}_T\left(t \mid \widehat{\sigma}(\boldsymbol{x})\widehat{\varepsilon}_i + \widehat{\mu}(\boldsymbol{x}),\boldsymbol{x}\right)\right)},\label{cd_se_est}\\
\widehat{\text{Sp}}^\mathbb{D}(\upsilon,t \mid \boldsymbol{x}) = & \frac{\int_{-\infty}^{\upsilon}\widehat{S}_T\left(t \mid y, \boldsymbol{x}\right)\text{d}\widehat{F}_Y\left(y \mid \boldsymbol{x} \right)}{\int_{-\infty}^{\infty}\widehat{S}_T\left(t \mid y, \boldsymbol{x}\right)\text{d}\widehat{F}_Y\left(y \mid \boldsymbol{x} \right)} = \frac{\int_{-\infty}^{\upsilon}\widehat{S}_T\left(t \mid y, \boldsymbol{x}\right)\text{d}\widehat{F}_{\varepsilon}\left(\frac{y - \widehat{\mu}(\boldsymbol{x})}{\widehat{\sigma}(\boldsymbol{x})}\right)}{\int_{-\infty}^{\infty}\widehat{S}_T\left(t \mid y, \boldsymbol{x}\right)\text{d}\widehat{F}_{\varepsilon}\left(\frac{y - \widehat{\mu}(\boldsymbol{x})}{\widehat{\sigma}(\boldsymbol{x})}\right)}\nonumber\\
= & \frac{\sum_{i=1}^{n}\widehat{S}_T\left(t \mid \widehat{\sigma}(\boldsymbol{x})\widehat{\varepsilon}_i + \widehat{\mu}(\boldsymbol{x}), \boldsymbol{x}\right)I\left(\widehat{\sigma}(\boldsymbol{x})\widehat{\varepsilon}_i + \widehat{\mu}(\boldsymbol{x}) \leq \upsilon \right)}{\sum_{i=1}^{n} \widehat{S}_T\left(t \mid \widehat{\sigma}(\boldsymbol{x})\widehat{\varepsilon}_i + \widehat{\mu}(\boldsymbol{x}), \boldsymbol{x}\right)}\label{cd_sp_est}.
\end{align}
We note that estimating $F_{\varepsilon}(\cdot)$ using the empirical distribution function of the standardised residuals allows the integrals in the first lines of Equations \eqref{cd_se_est} and \eqref{cd_sp_est} to be replaced with summations, thereby speeding up computations. Regarding the estimation of the ROC curve given in \eqref{ROC_curve}, for a given value of $p$, it is obtained by linear interpolation based the on Se and Sp estimates over a fine grid of threshold values for $v$. Finally, the AUC in \eqref{AUC} is estimated by approximating the integral using the composite Simpson's rule.
\subsubsection{\textbf{\textsf{Specification of Smooth Functions}}}\label{smooth_fun}
As noted earlier, under the model specifications given in Equations \eqref{hazard_gam} and \eqref{loc_scale_model_2}, smooth nonlinear effects of continuous variables can be incorporated using penalised regression splines. In this section, we briefly outline the main ideas behind penalised splines and refer readers to \cite{Wood2017} (and the references therein) for a detailed exposition. 

Penalised splines approximate nonlinear effects using a linear combination of known spline basis functions and adjust the log-likelihood (or residual sum of squares) by adding a penalty term to prevent overfitting. 
In the one-dimensional case, each nonlinear function is expressed as
\[
s\left(u\right) = \sum_{j=1}^{J}\theta_jB_j\left(u\right),
\]
where $B_j\left(\cdot\right)$, $j = 1, \ldots, J$, are known spline basis functions (such as cubic regression splines, our choice here, given their good performance, or B-splines)
and $\boldsymbol{\theta} = \left(\theta_1, \cdots, \theta_J\right)^{\top}$ is the vector of unknown regression coefficients. Smooth varying-coefficient terms, i.e., terms of the form $u_1s(u_2)$, where $u_1$ and $u_2$ are continuous covariates, can be specified analogously. For two-dimensional smooth functions (with higher-dimensional extensions following a similar approach), a tensor product of two marginal bases is used, i.e.,
\[
s\left(u_1, u_2\right) = \sum_{j_1=1}^{J_1}\sum_{j_2=1}^{J_2}\theta_{j_1j_2}B_{1j_1}\left(u_1\right)B_{2j_2}\left(u_2\right), \quad \boldsymbol{\theta} = \left(\theta_{11}, \theta_{12}, \cdots, \theta_{J_1J_2}\right)^{\top}.
\]
In all cases, estimating $s(\cdot)$ reduces to estimating $\boldsymbol{\theta}$, but the number of basis functions must be specified in advance. Penalised splines rely on setting $J$ (or $J_1$ and $J_2$ in the two-dimensional case) to a moderate or large value and then adding a penalty to the objective function to enforce smoothness. Typically, the penalty is based on the integral of the $r$th-order (partial) derivative of the function $s(\cdot)$. For the one- and two-dimensional cases, the penalty takes the following forms, respectively
\begin{align*}
\text{Pen}_{\text{1D}} = \lambda\boldsymbol{\theta}^{\top}\boldsymbol{P}\boldsymbol{\theta}\quad\text{and}\quad
\text{Pen}_{\text{2D}} = \lambda_1\boldsymbol{\theta}^{\top}\left(\boldsymbol{P}_1\otimes\boldsymbol{I}_{J_2}\right)\boldsymbol{\theta} + \lambda_2\boldsymbol{\theta}^{\top}\left(\boldsymbol{I}_{J_1}\otimes\boldsymbol{P}_2\right)\boldsymbol{\theta},
\end{align*}
where $\boldsymbol{I}_{N}$ is an identity matrix of order $N \times N$ and $\otimes$ denotes the Kronecker product. The specific form of the known penalty matrix $\boldsymbol{P}$ (or $\boldsymbol{P}_1$ and $\boldsymbol{P}_2$) depends on the chosen basis functions and the penalty order $r$. Following standard practice, we use a second-order penalty ($r = 2$), which is the default and most widely recommended choice in the literature \cite[see, e.g.,][]{Eilers1996, Wood2017}. Finally, $\lambda$ (or $\lambda_1$ and $\lambda_2$) is the smoothing parameter that controls the amount of smoothing, that is, the trade-off between fidelity to the data and smoothness of the function estimate. Various methods can be used to select appropriate values for the smoothing parameters, such as generalised cross-validation (GCV) or the Akaike information criterion (AIC). We use restricted maximum likelihood (REML), which exploits the connection between penalised splines and mixed-effects models. REML tends to yield smoothing parameter estimates with lower variability than alternative criteria, while avoiding the undersmoothing and multiple-minima issues often observed with AIC and GCV \citep{ReissOgden2009, Krivobokova2013, Wood2017}.

For brevity, the details above focus on a single smooth function. However, the extension to multiple smooth functions is straightforward, with the important consideration that, to ensure identifiability of models \eqref{hazard_gam} and \eqref{loc_scale_model_2}, all included smooth functions must be subject to sum-to-zero constraints \cite[][Sections 5.4.1 and 5.6.3]{Wood2017}. 

\subsubsection{\textbf{\textsf{Inference}}}
Inference is performed through a case-resampling bootstrap, i.e., our resampling procedure is based on sampling with replacement from the original dataset. For each bootstrap sample, the covariate-specific time-dependent ROC curve and the corresponding AUC are computed. The performance of this bootstrap scheme, when combined with the percentile method for constructing confidence intervals for the covariate-specific time-dependent AUC, is evaluated in the simulation study presented in Section~\ref{sim_main}.

\subsection{\textbf{\textsf{Software Implementation Details}}}\label{implementation}
As introduced in Section \ref{method}, our approach to estimating the covariate-specific cumulative-dynamic time-dependent ROC curve is based on: (i) fitting a penalised-spline-based Poisson GAM to estimate the conditional hazard function using the piecewise exponential (PAM) representation, and (ii) fitting two penalised-spline-based AMs to estimate the mean and variance functions of the model posit for the biomarker given the covariates. To facilitate the application of our method in practice, we have developed the publicly available \texttt{R} package \texttt{CondTimeROC}, which uses \texttt{mgcv} \citep{Wood2017} for model fitting and \texttt{pammtools} \citep{bender2018PAM_art} for PAM-specific data preprocessing. For completeness, \texttt{CondTimeROC} also implements the methods proposed by~\cite{xiao08} and ~\cite{MX16}. The package is available at \url{https://github.com/mxrodriguezUVIGO/CondTimeROC}.

\section{\large{\textsf{Simulation Study}}}\label{sim_main}
A simulation study is carried out to investigate the empirical performance of the penalised-spline based estimator introduced in Section \ref{method}.
For comparison purposes, we also include in our study the semiparametric approach proposed by~\cite{xiao08} and the fully (smoothed) nonparametric proposal by~\cite{MX16}. As the latter accommodates only one continuous covariate, simulations are restricted to that setting.  Simulations are performed in \texttt{R} \citep{R24} using the \texttt{CondTimeROC} package, with plots generated via \texttt{ggplot2} \citep{Wickham16}.

\subsection{\textsf{Simulation Scenarios and Implementation Details}}\label{scenarios}
We consider three scenarios: (I) the proportional hazards assumption holds and the effects for both the biomarker and the covariate are linear, (II)  the proportional hazards assumption holds and the effects for both the biomarker and the covariate are nonlinear, and (III) the proportional hazards assumption does not hold and the effects for both the biomarker and the covariate are nonlinear. Specifically, the scenarios are as follows
\begin{itemize}
\item \textbf{Scenario I}: Linear effects and proportional hazards
\begin{align*}
\lambda\left(t \mid y, x\right) & = \exp\left(2 + \log(t + 0.2) + y + 0.1x\right),\\
Y & = x + \varepsilon.
\end{align*}
\item \textbf{Scenario II}: Nonlinear effects and proportional hazards
\begin{align*}
\lambda\left(t \mid y, x\right) & = \exp\left(2 + \log(t + 0.2) + (y^3/20) + 0.5\sin(2(x + 1.5))\right),\\
Y & = 0.2(x+0.5)^2 + \left(0.6(x-1)^2 + 0.4\right)\varepsilon.
\end{align*}
\item \textbf{Scenario III}: Nonlinear effects and non-proportional hazards
\begin{align*}
\lambda\left(t \mid y, x\right) & = \exp\left((2 + \log(t + 0.2))(y^3/20) + 0.5\sin(2(x + 1.5))\right),\\
Y & = 0.2(x+0.5)^2 + \left(0.6(x-1)^2 + 0.4\right)\varepsilon.
\end{align*}
\end{itemize}
In all three scenarios, we consider $X\sim \mbox{N}\left(1,1\right)$ and $\varepsilon \sim \mbox{N}(0,1)$, and we further assume covariate-dependent censoring, where the censoring time $C$ is generated from an exponential distribution with rate $\frac{1}{a + b|X|}$, truncated at $20$. The values $a$ and $b$ are chosen to achieve an unconditional censoring rate, $\Pr\left(T > C\right)$, of approximately $50$\% in each scenario. For each scenario and sample size $n \in \{300, 600\}$, we generate 500 datasets.

To evaluate the performance of the estimators, we measure the discrepancy, at a fixed time $t$, between the estimated and true covariate-specific cumulative-dynamic time-dependent ROC curves using the  empirical root mean squared error (ERMSE) 
\[
\mbox{ERMSE}\left(t\right)= \sqrt{\frac{1}{n_X n_T}\sum_{j = 1}^{n_{X}}\sum_{k = 1}^{n_{T}}\left(\widehat{\mbox{ROC}}^{\mathbb{C}/\mathbb{D}}\left(p_k, t \mid x_j\right) - \text{ROC}^{\mathbb{C}/\mathbb{D}}\left(p_k, t \mid x_j\right)\right)^2},
\]
with $x_j = -1 + 4\frac{j-1}{n_X - 1}$, $j = 1,\ldots, n_X$, $p_k = \frac{k-1}{n_P - 1}$, $k = 1,\ldots, n_P$,  $n_X = 50$, and $n_P = 101$. For the covariate-specific cumulative Se and dynamic Sp, the  ERMSE is computed in a similar fashion but over a sequence of threshold values with $\upsilon_l = -3 + 8\frac{l-1}{n_\upsilon - 1}$, $l = 1,\ldots, n_\upsilon$ and $n_\upsilon = 200$. As for the covariate-specific time-dependent AUC, the performance of the estimators is evaluated in terms of bias. In all cases, the values of $t$ considered are the quartiles of the observed times: ($0.015$, $0.05$, $0.1$), ($0.04$, $0.12$, $0.24$), and ($0.09$, $0.26$, $0.58$) for Scenarios I, II, and III, respectively.

\subsubsection{\textbf{\textsf{Data Generating Procedure}}}\label{dg_procedure}
To generate the covariates, biomarker outcomes, and the survival times based on the models for the conditional hazard function described above, the following procedure is followed \cite[see, e.g.,][for more details]{Crowther2013}.

\noindent For $i = 1,\ldots,n$ 
\begin{description}
\item[Step 1.] Generate $x_i \overset{\text{iid}}\sim \text{N}\left(1,1\right)$, $y_i \overset{\text{ind}}\sim \text{N}\left(x_i, 1\right)$ (Scenario I) or $y_i \overset{\text{ind}}\sim \text{N}\left(0.2(x_i+0.5)^2, 0.6(x_i-1)^2 + 0.4\right)$ (Scenarios II and III), and $c_i = \max\{\tilde{c}_i, 20\}$ with $\tilde{c}_i \overset{\text{ind}}\sim \exp\left (\frac{1}{a + b|x_i|}\right)$.  
\item[Step 2.] Generate $u_i \overset{\text{iid}}\sim \text{U}(0,1)$ and obtain $t_i$ such that $S^{-1}_T\left(u_i\mid y_i, x_i\right) = t_i$; that is, $t_i$ is the value that satisfies $S_T\left(t_i\mid y_i, x_i\right) - u_i = 0$. To that end, the $\texttt{R}$ function $\texttt{uniroot}$ is used. 
\end{description}
One additional step is required to implement the data-generating procedure described: computing $S_T\left(\cdot \mid y_i,  x_i\right)$. Once again, we use the equality in \eqref{S_lambda_rel} along with the composite Simpson's rule to approximate the integral.

\subsubsection{\textbf{\textsf{Implementation Details}}}\label{sim_implementation}
For the penalised spline-based estimator we propose in this paper and for all scenarios, the following models are considered
\begin{equation*}
\lambda\left(t \mid y, x\right) = \exp\left(\alpha_0 + s_0(t) + s_X(x) + s_Y(y) + s_{X,T}(x,t) + s_{Y,T}(y,t)) + s_{X,Y}(x,y)\right),
\end{equation*}
and
\begin{equation*}
Y = \beta_0 + s_{\mu}\left(x\right) + \exp\left(\beta_1 + s_{\sigma}\left(x\right)\right)\varepsilon.
\end{equation*}
To approximate the one- and two-dimensional smooth functions involved in the model for the conditional hazard function, $\lambda\left(t \mid y, x\right)$, we use cubic regression splines and their tensor products with marginal bases of dimension $J = J_1 = J_2 = 8$ and second-order penalties. For the model for the biomarker $Y$, we also use cubic regression splines and a second-order penalty but with $J = 13$ basis functions. The difference in the $J$ values between the two models is solely due to computational reasons. In all cases, REML is used to select the smoothing parameters. Note that, for all scenarios, the models considered, particularly those for the conditional hazard function, are more complex than would be strictly necessary based on the data-generating mechanisms. Our aim is not to promote overly complex specifications, but rather to evaluate our approach in \textit{a priori} unfavourable conditions. Additionally, we evaluate the double penalty approach for model building and selection, as described in \cite{Marra11} and applied to hazard regression by \cite{Girondo2013}. This approach--used for both the model for the conditional hazard function and the location-scale model for the biomarker--penalises the otherwise unpenalised part of each smooth term and allows specific smooth functions to be entirely excluded from the model (i.e., estimated as zero) if deemed unnecessary \cite[for details, see][]{Marra11, Girondo2013}. Regarding the data augmentation strategy associated with the piecewise exponential approach, we consider the observed uncensored times as the break points for the follow-up period (see \eqref{break_points}). 

For the proposal by \cite{xiao08}, the proportional hazards model includes linear effects for both the covariate and the biomarker and a linear model is assumed for the biomarker. Finally, for the smoothed nonparametric estimator by \cite{MX16}, all bandwidths are selected using the least-squares cross-validation method proposed by~\cite{Li13} and a Gaussian kernel is employed.

\subsection{\textsf{Results}}
\begin{figure}
\centering
\includegraphics[width=12cm]{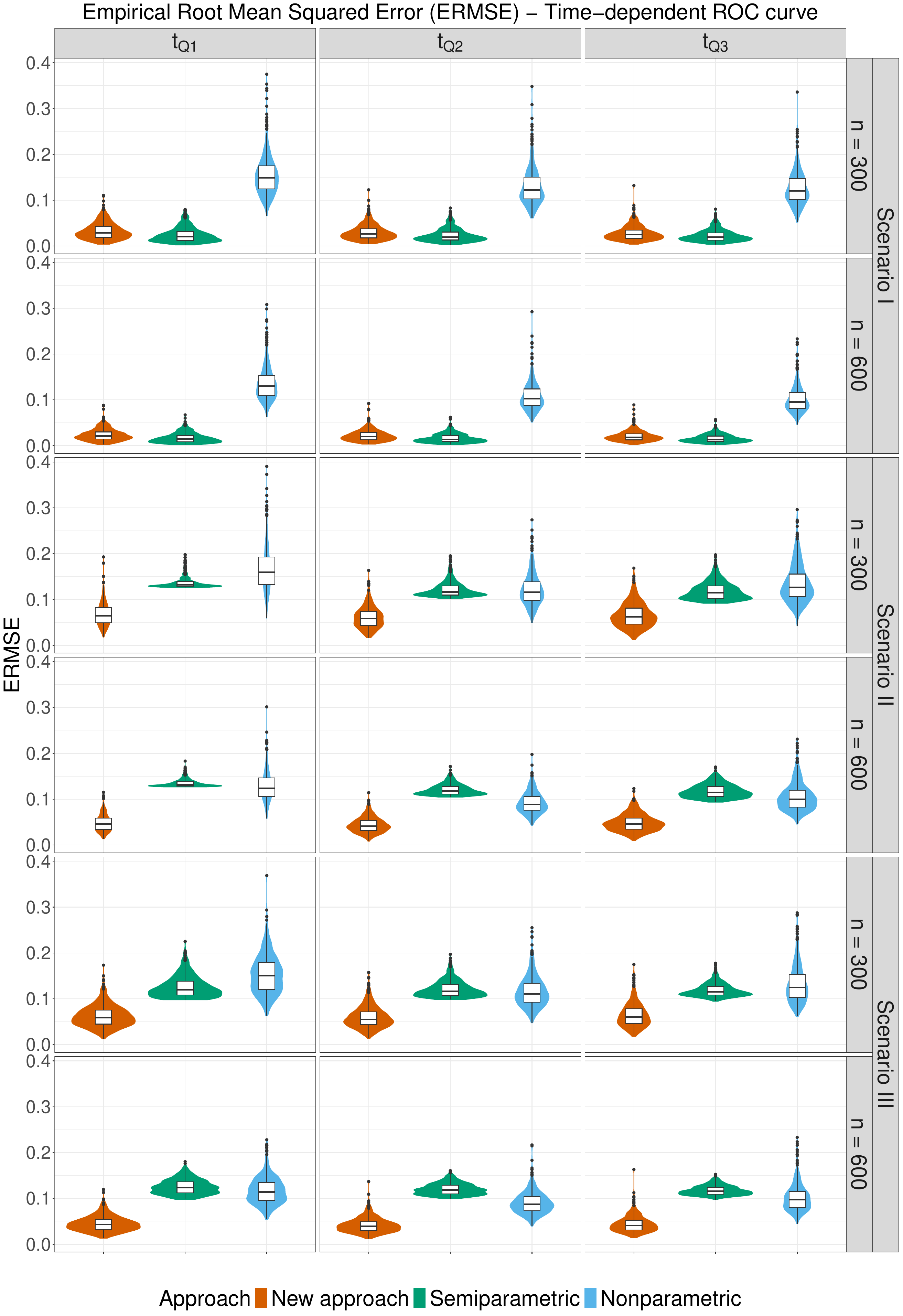}
\caption {Violin plot of the empirical root mean squared error for the covariate-specific cumulative-dynamic time-dependent ROC curve (over a sequence of covariate values) at the quartiles of the observed times ($t_{\text{Q1}}$, $t_{\text{Q2}}$ and $t_{\text{Q3}}$). `New approach' refers to our proposal, `Semiparametric' refers to the approach of~\cite{xiao08}, and `Nonparametric' to the smoothed nonparametric method of~\cite{MX16}.}
\label{ROC_RMSE_sim}
\end{figure}

\begin{figure}
\centering
\includegraphics[width=12cm]{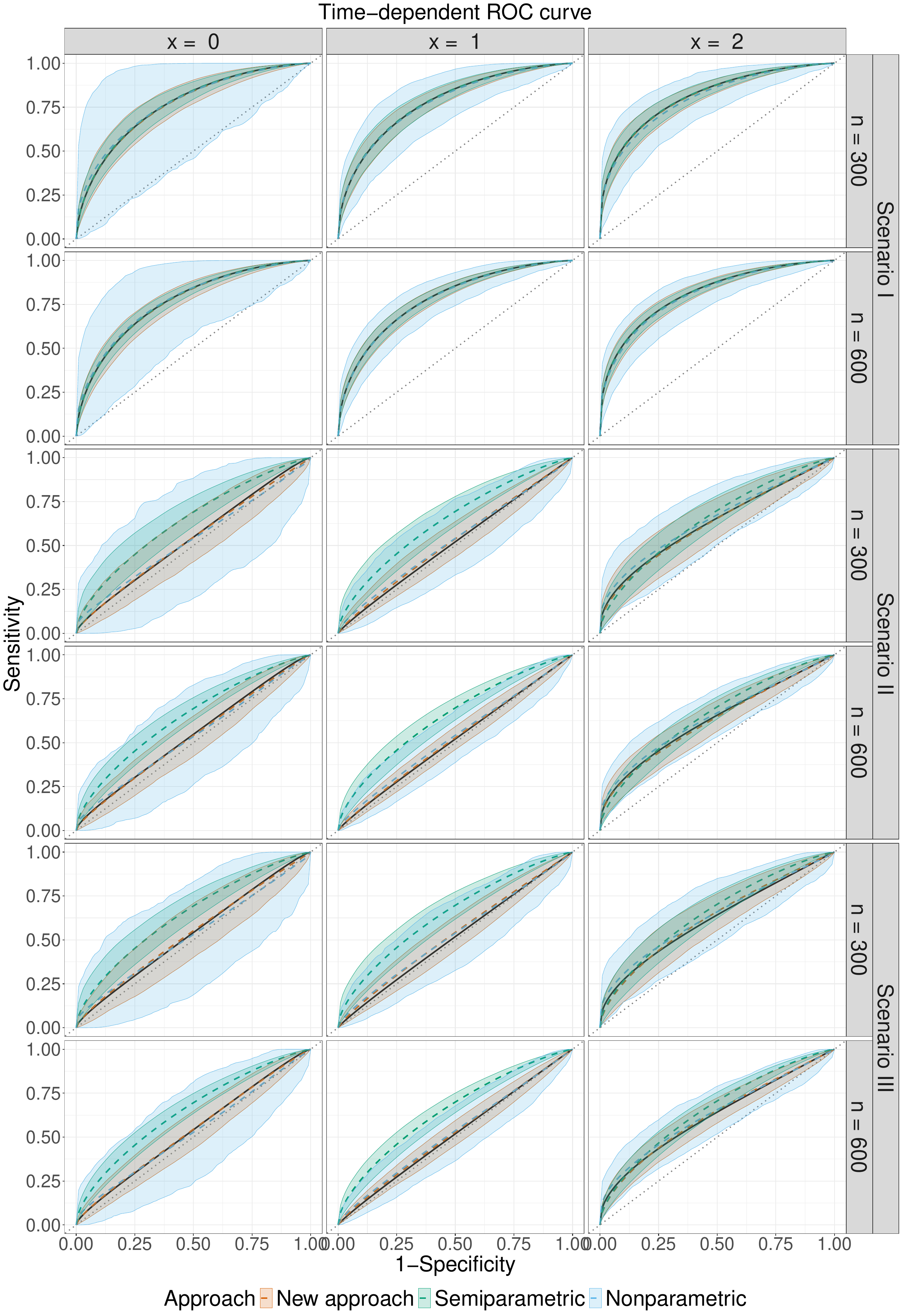}
\caption {True covariate-specific cumulative-dynamic time-dependent ROC curve (solid black line) versus the average of the estimated ROC curves (dotted coloured lines), evaluated at the median observed time ($t_{\text{Q2}}$) and for $x \in \{0, 1, 2\}$. The shaded areas are bands constructed using the pointwise 2.5\% and 97.5\% percentiles across simulations. `New approach' refers to our proposal, `Semiparametric' refers to the approach of~\cite{xiao08}, and `Nonparametric' to the smoothed nonparametric method of~\cite{MX16}.}
\label{ROC_Curves_sim}
\end{figure}

\begin{figure}
\centering
\includegraphics[width=12cm]{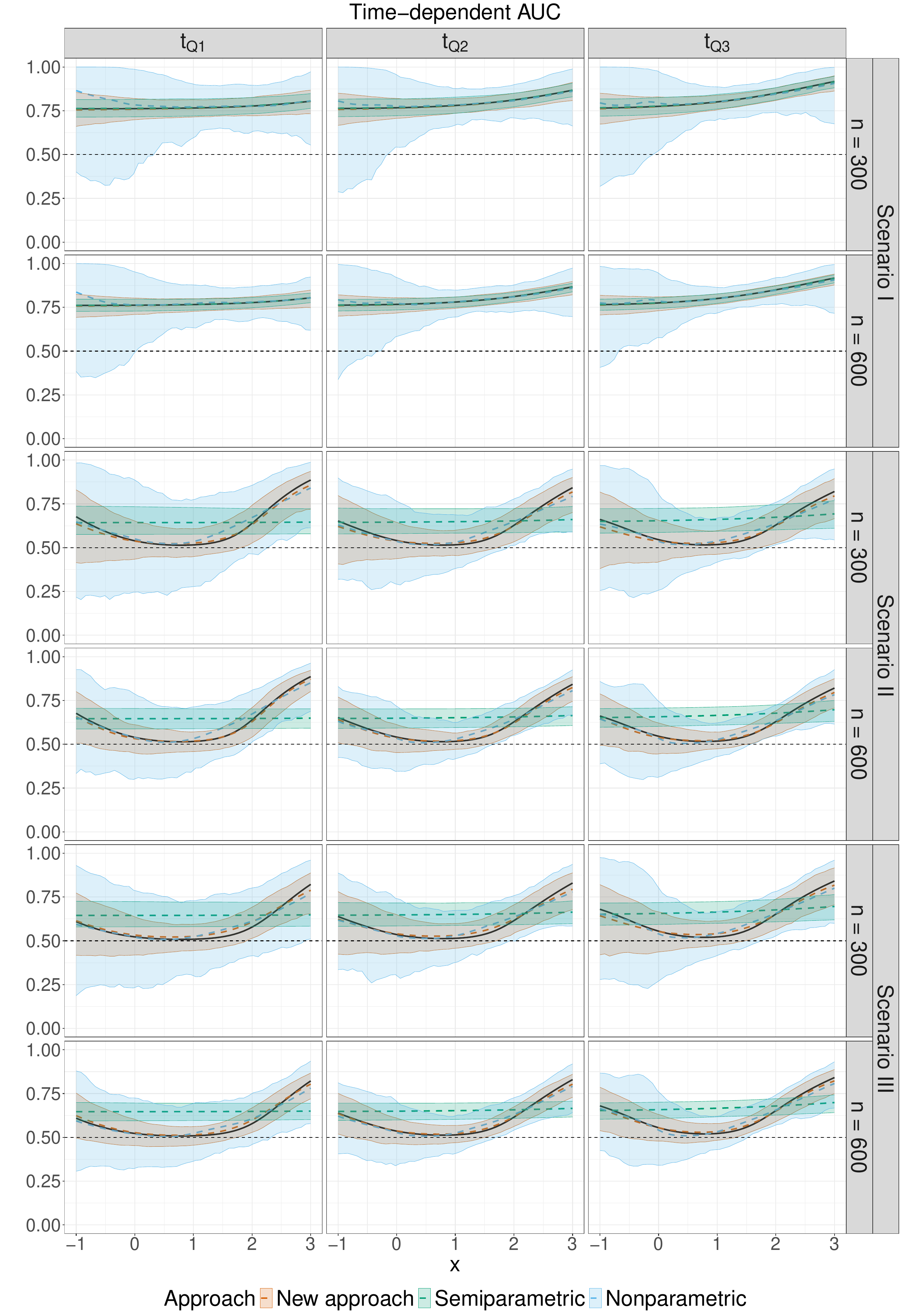}
\caption {True covariate-specific cumulative-dynamic time-dependent AUC (solid black line) versus the average of the estimated AUCs (dotted coloured lines), evaluated at the quartiles of the observed times ($t_{\text{Q1}}$, $t_{\text{Q2}}$, and $t_{\text{Q3}}$). The shaded areas are bands constructed using the pointwise 2.5\% and 97.5\% percentiles across
simulations. `New approach' refers to our proposal, `Semiparametric' refers to the approach of~\cite{xiao08}, and `Nonparametric' to the smoothed nonparametric method of~\cite{MX16}.}
\label{AUC_sim}
\end{figure}

For brevity, the majority of graphical results are provided in \ref{supp:simulation} and we focus here on the main findings. Figure \ref{ROC_RMSE_sim} shows violin plots of the ERMSE for the covariate-specific time-dependent ROC curve, averaged over the sequence of covariate values, across the 500 simulated datasets, for each scenario, sample size, approach, and time point considered. To better evaluate the performance of the proposed method and its competitors, we present Figures \ref{ROC_Curves_sim} and \ref{AUC_sim}. Figure \ref{ROC_Curves_sim} shows the average of the estimated ROC curves, along with the 2.5\% and 97.5\%  pointwise simulation quantiles, evaluated at the median observed time ($t_{\text{Q2}}$) and for $x \in \{0, 1, 2\}$, representing the 15\%, 50\%, and 85\% quantiles of the covariate distribution, respectively. Results for the first ($t_{\text{Q1}}$) and third ($t_{\text{Q3}}$) quartiles of the observed times are provided in Web Figures \ref{ROC_Curves_sim_q1} and \ref{ROC_Curves_sim_q3}. Figure \ref{AUC_sim}, displays the average and simulation quantiles of the estimated covariate-specific time-dependent AUC across different covariate values and for all time points considered.

In Scenario I, the best-performing method is that of \cite{xiao08}. This result is expected, as in this scenario, the model by \cite{xiao08} is correctly specified, and the added flexibility of our proposal (nonlinear effects and non-proportional hazards) is not required. However, the results show that our approach remains very competitive, performing well even when the extra flexibility is unnecessary. It exhibits negligible bias, with only a slight increase in variability compared to the method of \citet{xiao08}. As expected, and consistently across all methods and scenarios, the variability decreases as the sample size increases. For Scenarios II and III, which involve nonlinear effects of the covariate and the biomarker, non-proportional hazards, or both, our method consistently outperforms its competitors, successfully reconstructing the true ROC curves and AUCs with substantially lower variability than the nonparametric approach by \cite{MX16}. In contrast, the model proposed by \cite{xiao08} exhibits systematic bias, resulting in unreliable estimates. Notably, our proposed approach aligns more closely with the data-generating mechanisms across all scenarios, unlike the method of \cite{MX16}. This, combined with the fact that the approach is fully nonparametric, may explain the high variability observed for this method. Although the proposal exhibits low bias and, on average, recovers both the ROC curve and AUC well across all scenarios, the variability remains substantial. This translates to the ERMSE, with this method generally performing the worst, particularly for small sample sizes, even in scenarios where the model by \cite{xiao08} is biased. Additional results are presented in \ref{supp:simulation_main_document}. In particular, we report the ERMSE for the covariate-specific cumulative Se and dynamic Sp, as well as the violin plot of the  AUC bias. Overall, the results are consistent with those described earlier.

In addition to the previous results, we evaluate the performance of our approach under two alternative configurations. In the first case, we omit the use of the double-penalty approach for model building/selection, while in the second case, we divide the follow-up period into 30 equally spaced intervals for the PAM representation used to estimate the model for the conditional hazard function. Results are provided in \ref{supp:simulation:w_wo_model_selection} and \ref{supp:simulation:diff_timepoints}, respectively. As observed, omitting the double-penalty slightly increases the variability of the estimates, but the method still performs well, even under unnecessarily complex model specifications. Regarding the division of the follow-up period, although using 30 equally spaced intervals reduces the size of the resulting data and leads to faster estimation times compared to using uncensored times as break points, this approach somewhat increases variance and bias, particularly in the most complex scenario (Scenario III). While more suitable divisions could be explored, this is beyond the scope of the present work.

Finally, we also evaluate the performance of the case-resampling bootstrap percentile method for constructing confidence intervals for the covariate-specific time-dependent AUC. Results based on 500 replicates are provided in Section~\ref{supp:simulation:coverage_width}. As shown, the bootstrap procedure yields coverage probabilities close to the nominal level (0.95); however, it tends to be slightly conservative, which likely results in wider confidence intervals.

\section{\large{\textsf{Real Data Application}}}\label{app_main}
The prognosis for acute coronary syndrome (ACS) survivors is influenced by a complex interplay of factors, including age, comorbid conditions, the extent of coronary artery disease, arrhythmia risk, and, most importantly, the degree of left ventricular systolic dysfunction, measured by the left ventricular ejection fraction (LVEF). Post-ACS patients with an LVEF below 40\% have the highest mortality rates following the initial event. The Global Registry of Acute Coronary Events (GRACE) risk score is the most widely used and extensively validated tool for predicting both in-hospital and post-discharge mortality in patients with ACS \citep{Eagle04}, and it is endorsed by multiple international clinical guidelines as a core component of patient risk stratification. However, although LVEF is a well-established prognostic indicator of mortality following ACS, it is not included in the GRACE score due to missing data in the original derivation cohort \citep{Steg12}. This omission raises concerns that GRACE-based risk estimates may be confounded by unaccounted LVEF effects. The work by \cite{MX16} explored this issue by assessing the impact of LVEF on the prognostic performance of the GRACE score using the covariate-specific cumulative-dynamic time-dependent ROC curve. Their findings showed that, while the GRACE score predicts mortality adequately in patients with LVEF $\geq 40$\%, its predictive ability is reduced in those with LVEF below 40\%, highlighting the need for additional clinical variables or biomarkers to guide risk stratification in this high-risk subgroup. We revisit this clinical question and apply the method proposed in this paper to assess whether incorporating LVEF as a covariate affects the time-dependent discriminatory ability of the GRACE risk score. This analysis serves to both evaluate the robustness of the earlier findings and demonstrate the added value of our semiparametric approach, which yields reduced variability and enables a more precise assessment of subgroup-specific prognostic performance. Before proceeding, it is worth emphasizing that our goal is to investigate how the temporal discriminatory ability of the GRACE risk score varies across groups of individuals defined by their LVEF values. This is a valid and conceptually distinct objective from assessing how a composite prediction model, formed by combining the GRACE score and LVEF, performs. The latter approach evaluates how much additional predictive information LVEF contributes when added to the score, whereas our aim is to examine whether, and how, the discriminatory ability of the GRACE score itself, which is widely used in practice, depends on LVEF. This covariate-specific perspective provides a clearer understanding of the contexts in which an existing risk score retains its prognostic value and those in which its performance may diminish, without modifying the underlying model or incorporating new predictors.

\subsection{\textsf{Data Source}}\label{app_data}
The study included 3488 consecutive patients admitted for ACS between November 2003 and January 2011 at Hospital Cl\'inico Universitario de Santiago de Compostela (Spain), all of whom survived their hospitalisation. After discharge, patients were followed through a specialised coronary artery disease clinic and primary care. A structured follow-up was conducted by reviewing electronic medical records to track all-cause mortality until December 2011. Patients lost to follow-up were censored at the date of their last recorded clinical encounter. The median follow-up duration from hospital discharge was 37.80 months (interquartile range: 20.90--60.97 months), with a censoring rate of approximately 81\%. The mean (standard deviation) GRACE score and LVEF were 119 (35.0) and 55.8 (10.9), respectively, with 14\% of patients having an LVEF below 40\%.

\subsection{\textsf{Data Analysis and Results}}\label{app_results}
To estimate the covariate-specific cumulative-dynamic time-dependent ROC curve with our proposed method, we consider the following models
\begin{align*}
\lambda\left(t \mid \text{GRACE}, \text{LVEF}\right) & = \exp(\alpha_0 + s_0(t) + s_\text{LVEF}(\text{LVEF}) + s_\text{GRACE}(\text{GRACE}) + \\
& ~~~~~~~~~~\text{LVEF}s_{\text{LVEF},\text{GRACE}}(\text{GRACE}) + ts_{\text{GRACE},T}(\text{GRACE})),
\end{align*}
and
\begin{equation*}
\text{GRACE} = \beta_0 + s_{\mu}\left(\text{LVEF}\right) + \exp\left(\beta_1 + s_{\sigma}\left(\text{LVEF}\right)\right)\varepsilon,
\end{equation*}
where model selection for the hazard function was performed using the Akaike Information Criterion. We used cubic regression spline bases of dimension $J = 13$ for the one-dimensional smooth (varying-coefficient) functions involved in both the model for the hazard, $\lambda\left(t \mid \text{GRACE}, \text{LVEF}\right)$, and the GRACE score. Second-order penalties were considered, and smoothing parameters were selected using REML. As in the simulation study, we considered the observed uncensored times as the break points for the follow-up period in the piecewise exponential approach. With this configuration, the hazard model involves more than one million observations, and estimation required 16 seconds on a workstation equipped with an Intel$^\circledR$ Core\texttrademark\ i9-14900 processor (32 cores), 64 GiB RAM, and Ubuntu 22.04 LTS. Web Figures \ref{hazard_model_effects} and \ref{loc_scale_model_effects} show the estimated effects on the hazard and on the GRACE score, respectively. Among the terms in the hazard model, only the main effect of time, $s_0(\cdot)$, and the varying-coefficient function $s_{\text{GRACE},\text{T}}(\cdot)$ exhibit nonlinear behaviour. In contrast, LVEF shows a pronounced nonlinear effect on both the mean and variance of the GRACE score.

\begin{figure}
\centering
\includegraphics[width=17cm]{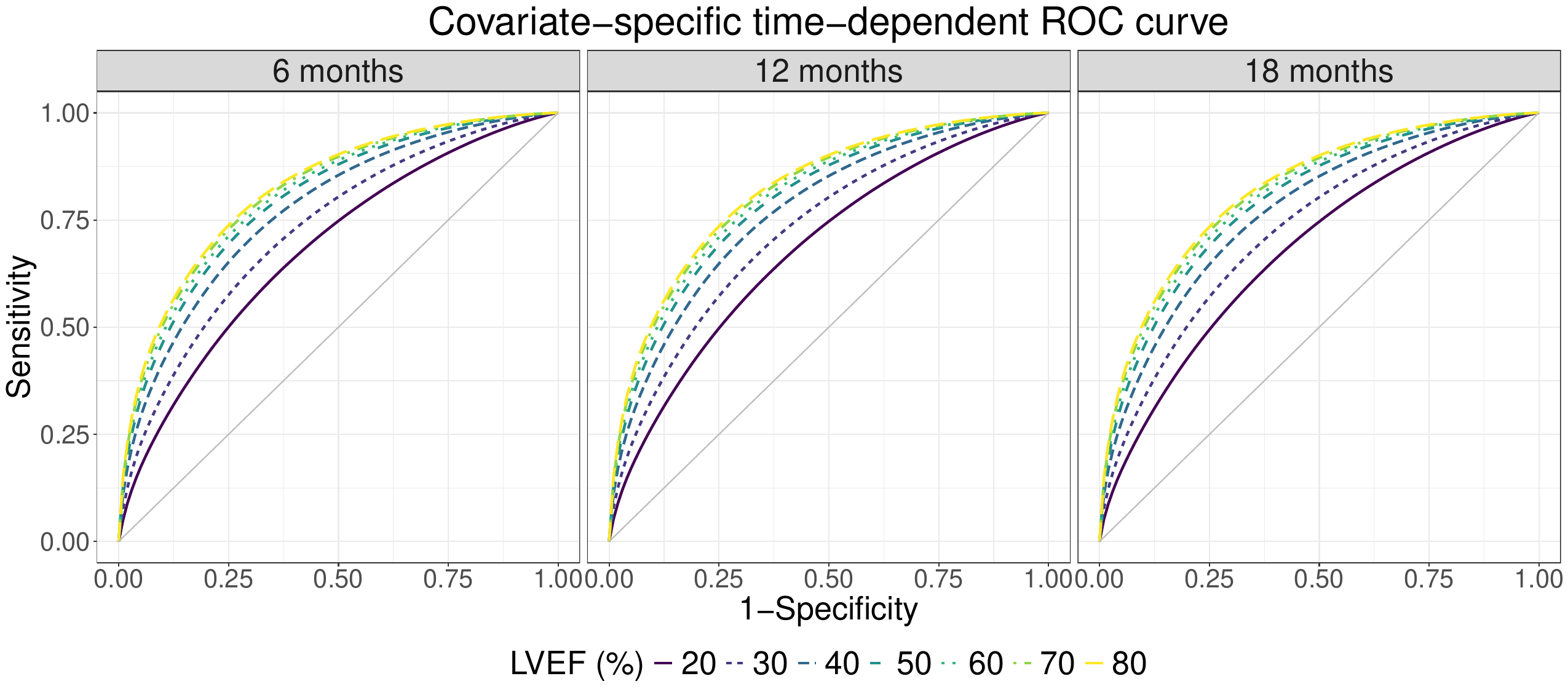}
\caption {Estimated covariate-specific cumulative-dynamic time-dependent ROC curve of the GRACE score, adjusted for LVEF(\%), evaluated at $t = 6, 12$ and $18$ months post-discharge. Results are shown for the method proposed in this paper.}
\label{ROC_Real_Data}
\end{figure}

\begin{figure}
\centering
\includegraphics[width=17cm]{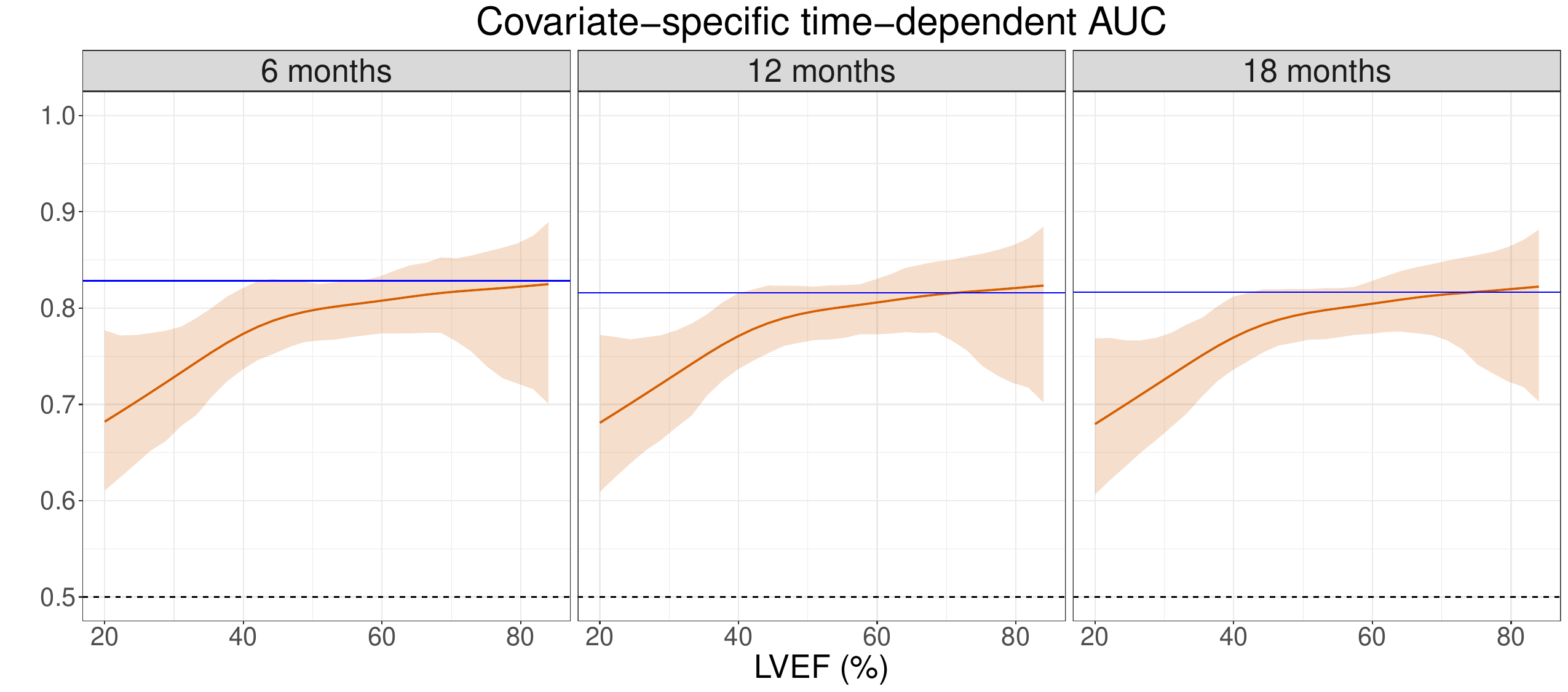}
\caption {Estimated covariate-specific cumulative-dynamic time-dependent AUC of the GRACE score, adjusted for LVEF(\%), with 95\% pointwise bootstrap confidence bands, evaluated at $t = 6, 12$ and $18$ months post-discharge. Results are shown for the method proposed in this paper. The blue lines correspond to the estimated pooled cumulative-dynamic time-dependent AUCs using the approach of \cite{Uno2007}.}
\label{AUC_Real_Data_new_approach}
\end{figure}

Figure \ref{ROC_Real_Data} shows the estimated covariate-specific cumulative-dynamic time-dependent ROC curves for the GRACE score, calculated at different LVEF values and at $t = 6$, $12$, and $18$ months post-discharge. For clarity, results are shown for only seven distinct LVEF values. As observed, at all three evaluated time points, the discriminatory ability of the GRACE score varies with LVEF, with higher LVEF values being associated with improved performance. To complement these findings, the covariate-specific time-dependent AUCs were also computed, with confidence intervals obtained using the bootstrap percentile method with $500$ replicates. Figure \ref{AUC_Real_Data_new_approach} displays the results across a wider range of LVEF values. These results suggest a nonlinear effect of LVEF on the prognostic value of the GRACE score, with similar AUC values observed across the three time periods considered. Specifically, the prognostic value increases with LVEF up to approximately $40\%$, and beyond this value, the AUC stabilises around $0.80$. The figure also shows the estimated unconditional (i.e., pooled or marginal) cumulative-dynamic time-dependent AUCs, obtained using the approach proposed by \cite{Uno2007}, which do not adjust for LVEF. As can be seen, the pooled AUCs are generally larger than the covariate-specific AUCs across most LVEF values and time points. These results would therefore lead to overly optimistic conclusions about the prognostic value of the GRACE score for predicting post-discharge mortality in ACS patients.

Finally, for completeness, we also analysed the data using the semiparametric approach of \cite{xiao08} and the smoothed nonparametric approach of \cite{MX16}. Results are shown in Web Figure \ref{AUC_Real_Data_all_supp}. For the method of \cite{xiao08}, we considered two specifications of the proportional hazards model's linear predictor: one including a linear interaction between LVEF and the GRACE score, and one without, as originally specified in \cite{MX16}. As shown in the figure, omitting the interaction term yields an estimated AUC, evaluated at 6, 12, and 18 months post-discharge, that is nearly constant across the range of LVEF values and substantially different from the estimates obtained with the other two approaches. In contrast, including the interaction term allows the semiparametric model to capture the influence of LVEF on the prognostic value of the GRACE score, revealing an increasing trend with LVEF, although, as expected, limited to a linear effect. The results obtained using the smoothed nonparametric approach of \cite{MX16} are generally consistent with those from our proposal, supporting the presence of a nonlinear relationship. However, our estimator yields smoother estimates with narrower confidence intervals, allowing for more precise conclusions about the time-dependent, LVEF-specific prognostic performance of the GRACE risk score.
\section{\large{\textsf{Discussion}}}\label{sec:discussion}
In this work we have developed a penalised spline approach for estimating covariate-specific time-dependent ROC curves. Our method relies on two key components: (i) a flexible piecewise exponential additive model for the hazard of the event time given the biomarker and covariates, and (ii) a flexible location-scale regression model for the conditional distribution of the biomarker given the covariates. As demonstrated both in the simulation study and the data application, our approach strikes a balance between the semiparametric method of \cite{xiao08} and the fully nonparametric method of \cite{MX16}. Like the nonparametric method, it makes fewer assumptions, offering greater flexibility and broader applicability. Unlike the nonparametric method, it achieves lower variance, enabling more precise evaluation of  a biomarker's prognostic performance, while accounting for nonlinear effects and relaxing the proportional hazards assumption, both of which are often required in practice.
Our method is also computationally appealing and is implemented, along with the approaches of \cite{xiao08} and \cite{MX16}, in the publicly available \texttt{R} package \texttt{CondTimeROC}, accessible at \url{https://github.com/mxrodriguezUVIGO/CondTimeROC}. Despite its advantages, our approach remains model-based and thus relies on the adequacy of the modelling assumptions for both the hazard and biomarker distribution components.

Several extensions of our approach are possible. One such extension is its adaptation to competing risks, following \cite{Zheng2012}, which would allow for a more comprehensive assessment of biomarker performance in the presence of multiple event types. This is particularly relevant in clinical settings where various types of failure must be considered simultaneously. Another potential extension is the estimation of positive and negative predictive values, also based on \cite{Zheng2012}, which would provide additional insights into the practical utility of a biomarker in prognostic studies. Finally, although not explored here, the piecewise exponential additive model could be adapted to handle left-truncated data. This extension would be useful in studies where individuals are enrolled only after a specific event or time point; for example, when data are available only for individuals who have already survived a certain duration. Such an adaptation would broaden the applicability of our approach to a wider range of survival analysis scenarios.

The \texttt{R}-codes needed to reproduce the results of the simulation study are available at \url{https://github.com/mxrodriguezUVIGO/CondTimeROC-paper}.

\section*{\large{\textsf{Acknowledgments}}}
This research was supported by the Ramón y Cajal Grant RYC2019-027534-I and by Grant PID2023-148811NB-I00 funded by MICIU/AEI/10.13039/501100011033 and by ``ERDF/EU''.

\putbib[ctimeROC]
\end{bibunit}
\newpage

\begin{bibunit}[hapalike]

\setcounter{section}{0}
\setcounter{equation}{0}
\setcounter{figure}{0}
\setcounter{table}{0}
\renewcommand{\theequation}{\Alph{section}\arabic{equation}}
\addto\captionsenglish{\renewcommand{\figurename}{Web Figure}}
\addto\captionsenglish{\renewcommand{\tablename}{Web Table}}
\renewcommand\thesection{Web Appendix \Alph{section}}

\makeatletter
\renewcommand{\fnum@figure}{Web Figure \thefigure}
\renewcommand{\fnum@table}{Web Table \thetable}
\makeatother

\makeatletter
\def\verbatim@font{\linespread{1}\normalfont\ttfamily}
\makeatother

\title{Supplementary materials for ``Penalised spline estimation of covariate-specific time-dependent ROC curves''}
\author{\textsc{Mar\'ia Xos\'e Rodr\'iguez-\'Alvarez} and \textsc{Vanda~In\'acio}}
\date{}
\maketitle
This document contains supplementary materials for the paper ``Penalised spline estimation of covariate-specific time-dependent ROC curves''. \ref{supp:simulation} presents additional results for the simulation study of Section \ref{sim_main} of the main text, along with results from two additional studies: (1) omitting the use of the double-penalty approach for model building/selection; and (2) dividing the follow-up period into 30 equally spaced intervals for the piece-wise exponential additive (PAM) representation used to estimate the conditional hazard function. \ref{supp:app} presents additional results for the application discussed in Section \ref{app_main} of the main text.

\let\thefootnote\relax\footnotetext{Mar\'ia Xos\'e Rodr\'iguez-\'Alvarez, CITMaga and Departamento de Estat\'istica e Investigaci\'on Operativa, Universidade de Vigo, Vigo, Spain (\textit{mxrodriguez@uvigo.gal}). Vanda In\'acio, School of Mathematics, University of Edinburgh, Scotland, UK (\textit{vanda.inacio@ed.ac.uk})}

\section{Simulation Study}\label{supp:simulation}
\subsection{Extra Results}\label{supp:simulation_main_document}

\begin{figure}[H]
\centering
\includegraphics[width=12cm]{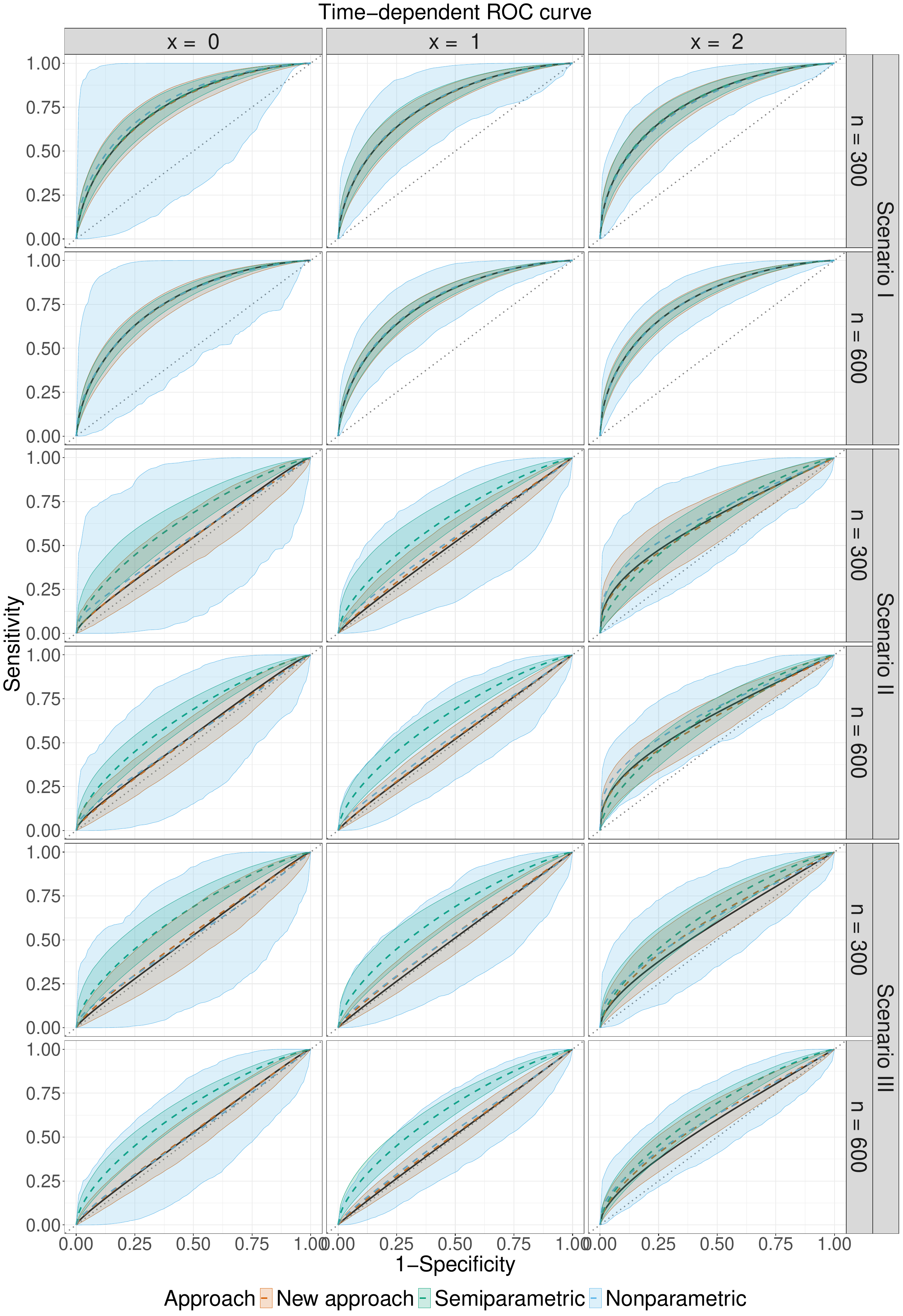}
\caption {True covariate-specific cumulative-dynamic time-dependent ROC curve (solid black line) versus the average of the estimated ROC curves (dotted coloured lines), evaluated at the first quartile of the observed time ($t_{\text{Q1}}$) and for $x \in \{0, 1, 2\}$. The shaded areas are bands constructed using the pointwise 2.5\% and 97.5\% quantiles across simulations. `New approach' refers to our proposal, `Semiparametric' refers to the approach of~\cite{xiao08}, and `Nonparametric' to the smoothed nonparametric method of~\cite{MX16}.}
\label{ROC_Curves_sim_q1}
\end{figure}
\begin{figure}[H]
\centering
\includegraphics[width=12cm]{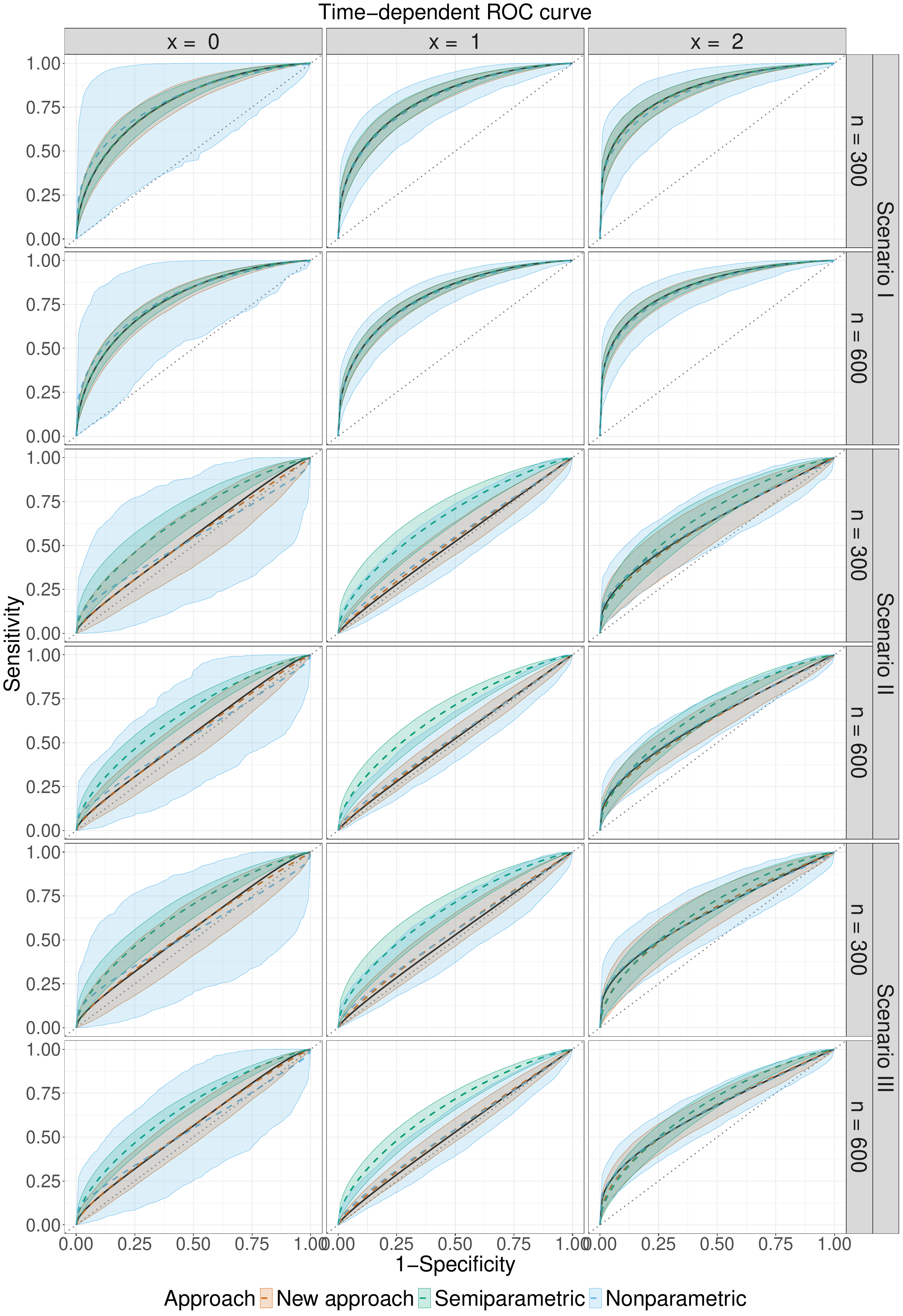}
\caption {True covariate-specific cumulative-dynamic time-dependent ROC curve (solid black line) versus the average of the estimated ROC curves (dotted coloured lines), evaluated at the third quartile of the observed time ($t_{\text{Q3}}$) and for $x \in \{0, 1, 2\}$. The shaded areas are bands constructed using the pointwise 2.5\% and 97.5\% quantiles across simulations.`New approach' refers to our proposal, `Semiparametric' refers to the approach of~\cite{xiao08}, and `Nonparametric' to the smoothed nonparametric method of~\cite{MX16}.}
\label{ROC_Curves_sim_q3}
\end{figure}
\begin{figure}[H]
\centering
\includegraphics[width=12cm]{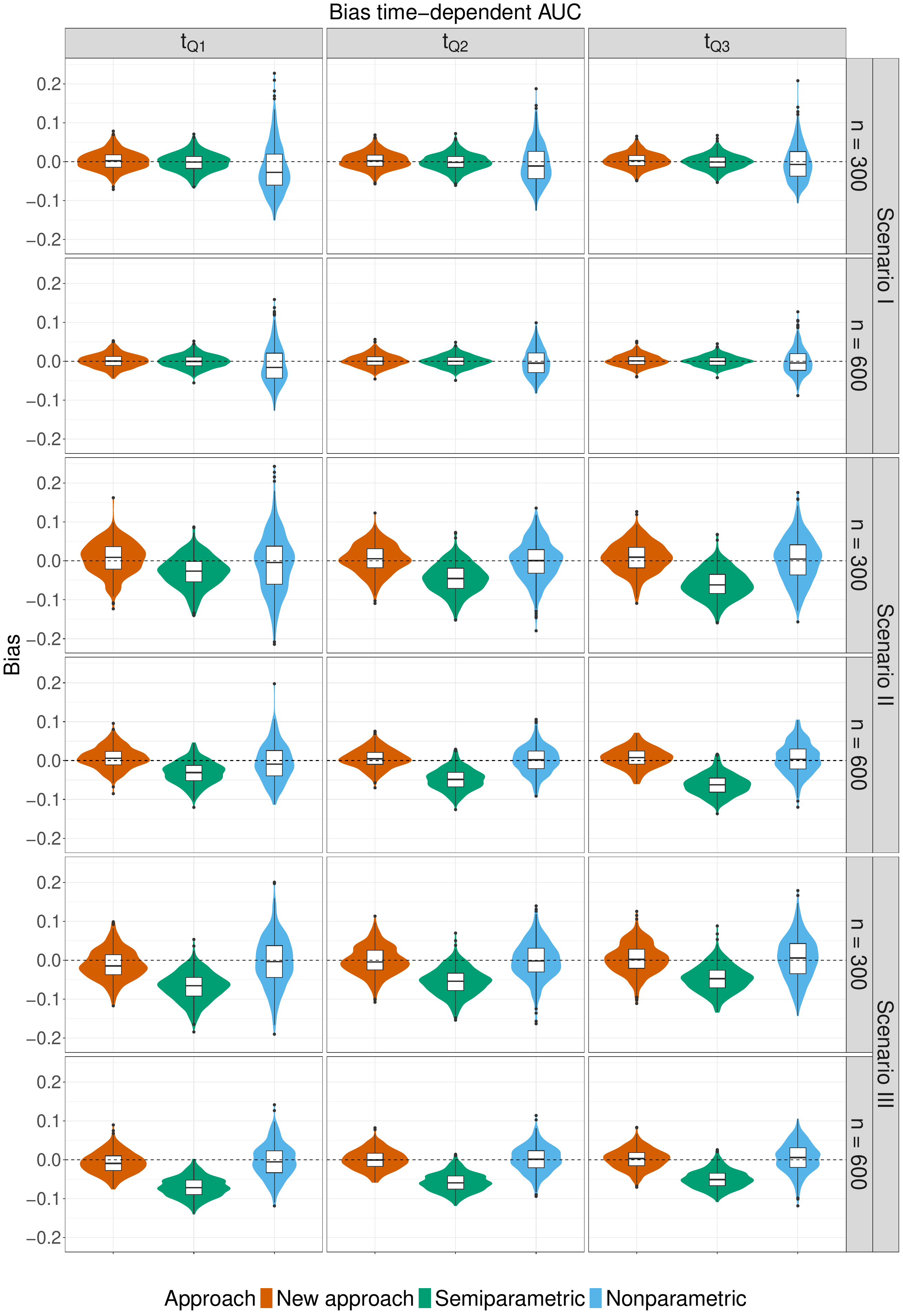}
\caption {Violin plot of the bias in the covariate-specific time-dependent cumulative AUC (over a sequence of covariate values) at the quartiles of the observed times ($t_{\text{Q1}}$, $t_{\text{Q2}}$, and $t_{\text{Q3}}$). `New approach' refers to our proposal, `Semiparametric' refers to the approach of~\cite{xiao08}, and `Nonparametric' to the smoothed nonparametric method of~\cite{MX16}.}
\label{Bias_AUC_sim}
\end{figure}
\begin{figure}[H]
\centering
\includegraphics[width=12cm]{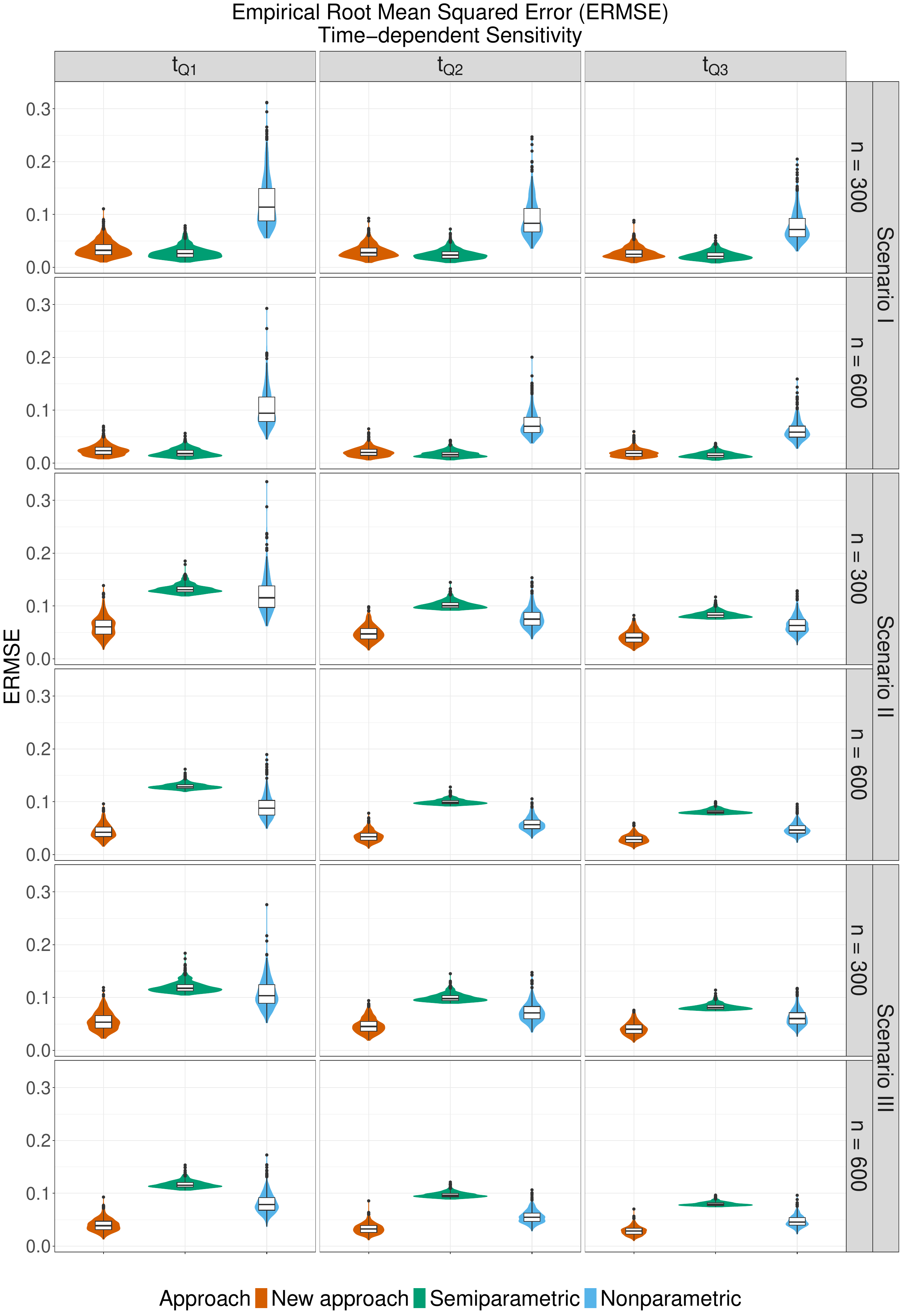}
\caption {Violin plot of the empirical root mean squared error for the covariate-specific time-dependent cumulative \textit{Sensitivity} (over a sequence of covariate values) evaluated at the quartiles of the observed times ($t_{\text{Q1}}$, $t_{\text{Q2}}$, and $t_{\text{Q3}}$). `New approach' refers to our proposal, `Semiparametric' refers to the approach of~\cite{xiao08}, and `Nonparametric' to the smoothed nonparametric method of~\cite{MX16}.}
\label{TPR_RMSE_sim}
\end{figure}
\begin{figure}[H]
\centering
\includegraphics[width=12cm]{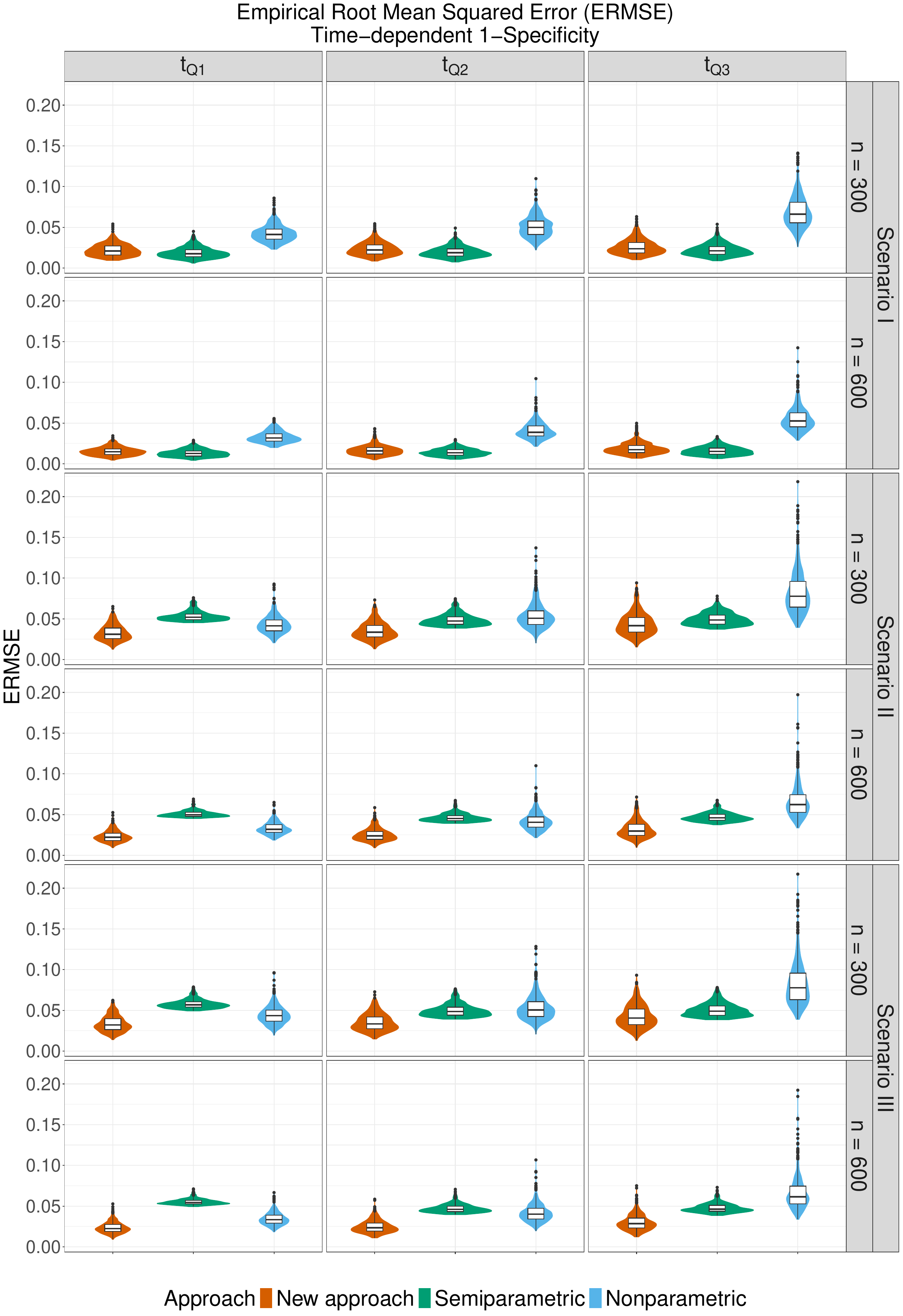}
\caption {Violin plot of the empirical root mean squared error for the covariate-specific time-dependent dynamic 1-\textit{Specificity} (over a sequence of covariate values) evaluated at the quartiles of the observed times ($t_{\text{Q1}}$, $t_{\text{Q2}}$, and $t_{\text{Q3}}$). `New approach' refers to our proposal, `Semiparametric' refers to the approach of~\cite{xiao08}, and `Nonparametric' to the smoothed nonparametric method of~\cite{MX16}.}
\label{FPR_RMSE_sim}
\end{figure}
\subsection{New Approach: Coverage and Widths for 95\% Bootstrapped Confidence Intervals}\label{supp:simulation:coverage_width} 
\begin{table}[H]
\caption{Coverage probabilities of the 95\% pointwise bootstrapped confidence interval for the covariate-specific time-dependent cumulative AUC evaluated at the quartiles of the observed times ($t_{\text{Q1}}$, $t_{\text{Q2}}$, and $t_{\text{Q3}}$) and at three covariate values ($x \in \{0, 1, 2\}$). Results are shown for the method proposed in this paper.}
\label{covegare_AUC}
\begin{center}
\small	
\begin{tabular}{ccccc}
& & & \multicolumn{2}{c}{Sample size}\\
Scenario & Time Value & Covariate Value & $300$ & $600$\\\hline
\multirow{9}{*}{I} &
\multirow{3}{*}{$t_{Q_1}$} & 0 & 98.2 & 98.4\\
 & & 1 & 97.8 & 98.0\\
 & & 2 & 96.4 & 98.4\\ \cline{2-5}
 & \multirow{3}{*}{$t_{Q_2}$} & 0 & 98.0 & 98.6\\
 & & 1 & 97.8 & 97.6\\
 & & 2 & 98.0 & 98.0\\ \cline{2-5}
 & \multirow{3}{*}{$t_{Q_3}$} & 0 & 97.8 & 98.8\\
 & & 1 & 98.2 & 97.8\\
 & & 2 & 98.6 & 97.6\\ \hline
\multirow{9}{*}{II} &
\multirow{3}{*}{$t_{Q_1}$} & 0 & 97.4 & 97.8\\
 & & 1 & 97.8 & 98.0\\
 & & 2 & 96.2 & 98.0\\ \cline{2-5}
 & \multirow{3}{*}{$t_{Q_2}$} & 0 & 98.0 & 97.8\\
 & & 1 & 97.6 & 98.2\\
 & & 2 & 96.8 & 98.4\\ \cline{2-5}
 & \multirow{3}{*}{$t_{Q_3}$} & 0 & 97.2 & 97.4\\
 & & 1 & 98.0 & 98.2\\
 & & 2 & 97.8 & 97.8\\ \hline
\multirow{9}{*}{III} &
\multirow{3}{*}{$t_{Q_1}$} & 0 & 96.8 & 96.4\\
 & & 1 & 97.8 & 97.6\\
 & & 2 & 96.2 & 96.8\\ \cline{2-5}
 & \multirow{3}{*}{$t_{Q_2}$} & 0 & 96.8 & 96.2\\
 & & 1 & 97.8 & 98.0\\
 & & 2 & 96.6 & 98.4\\ \cline{2-5}
 & \multirow{3}{*}{$t_{Q_3}$} & 0 & 97.6 & 96.2\\
 & & 1 & 96.6 & 97.8\\
 & & 2 & 98.2 & 97.6\\ \hline
\end{tabular}
\end{center}
\end{table}

\begin{figure}[H]
\centering
\includegraphics[width=12cm]{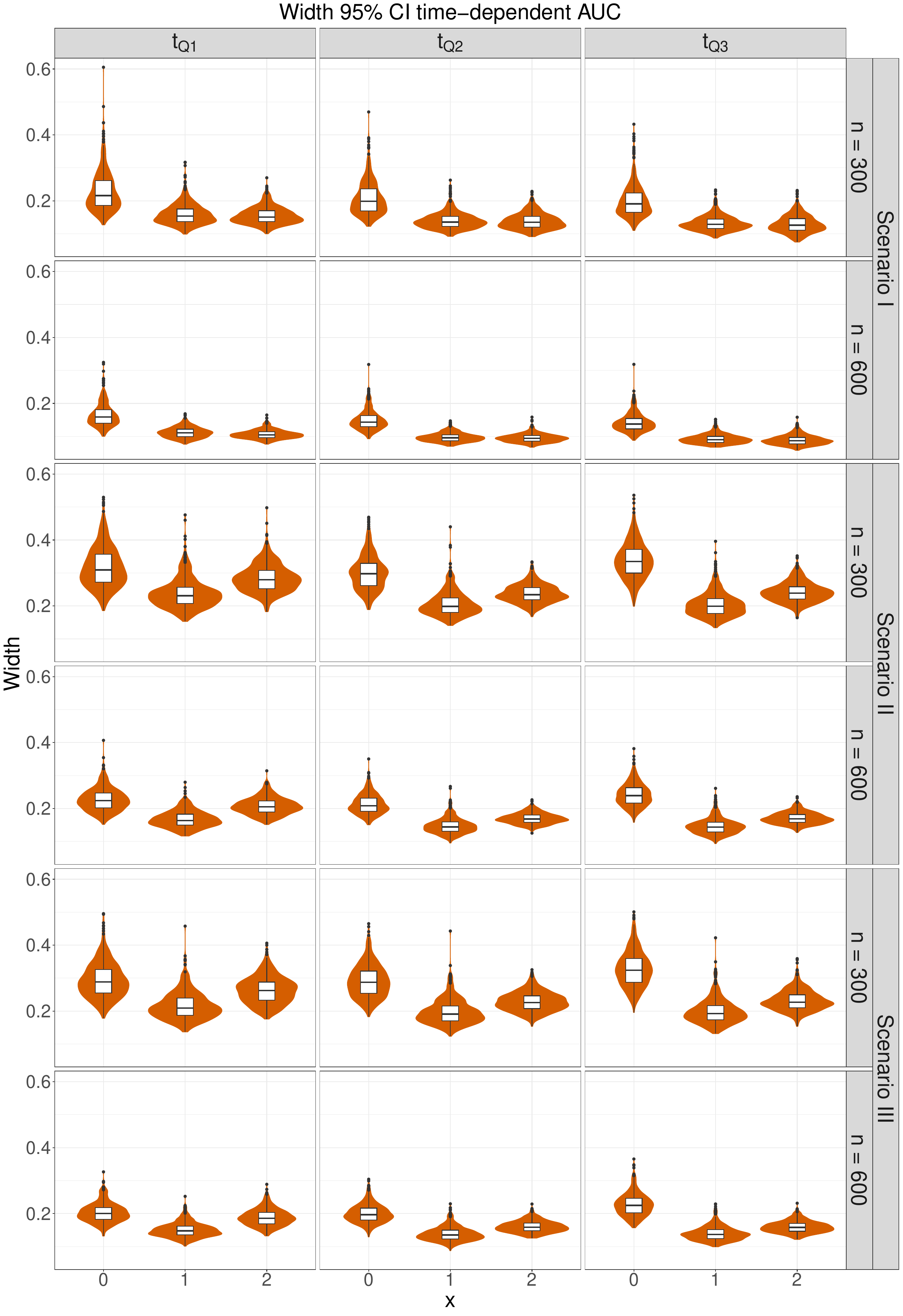}
\caption {Violin plot of the width of the 95\% pointwise bootstrapped confidence interval for the covariate-specific time-dependent cumulative AUC evaluated at the quartiles of the observed times ($t_{\text{Q1}}$, $t_{\text{Q2}}$, and $t_{\text{Q3}}$) and at three covariate values ($x \in \{0, 1, 2\}$). Results are shown for the method proposed in this paper.}
\label{coverage_width_sim}
\end{figure}

\subsection{New Approach: With and Without Model Selection}\label{supp:simulation:w_wo_model_selection}

\begin{figure}[H]
\centering
\includegraphics[width = 12cm]{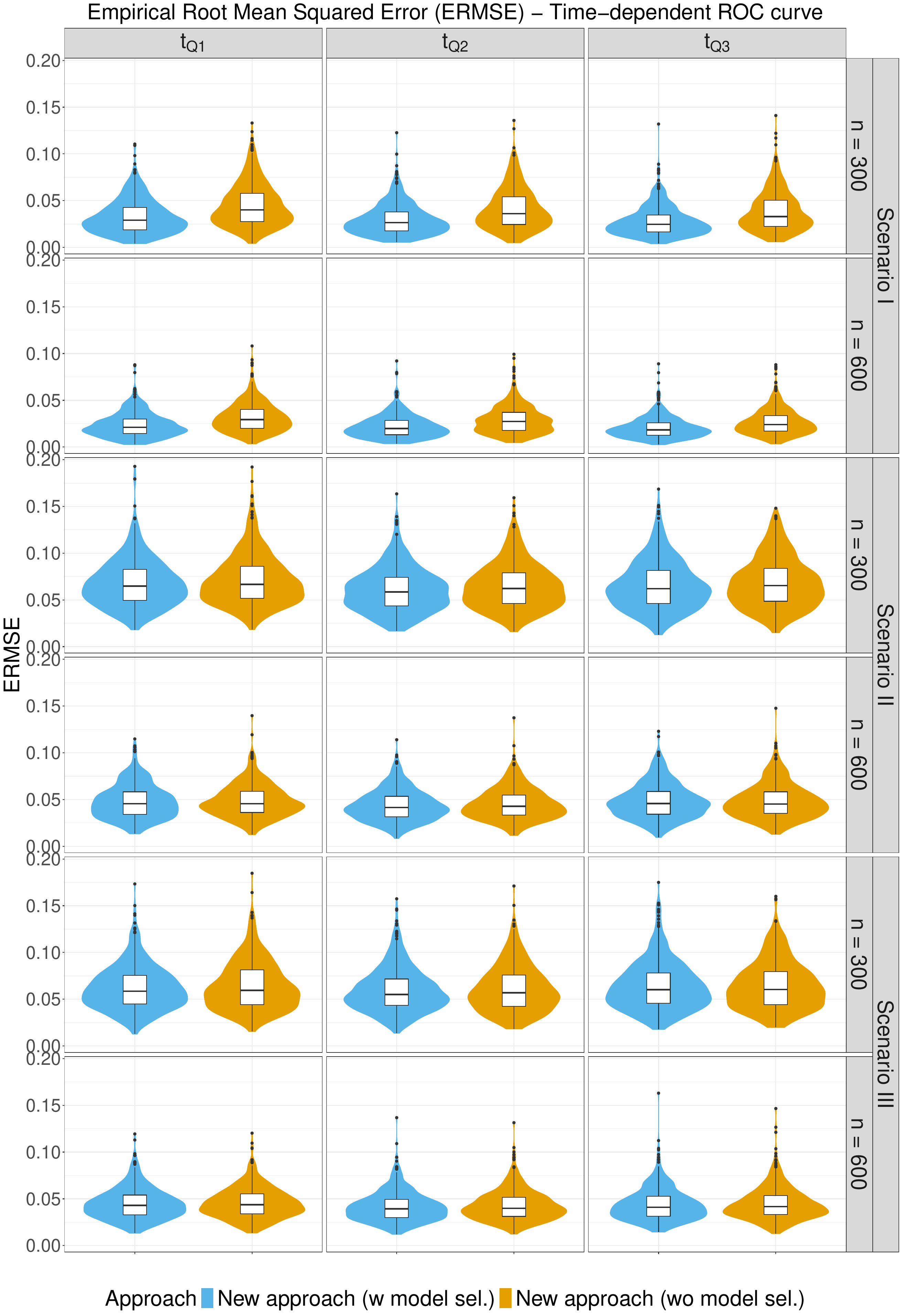}
\caption {Violin plot of the empirical root mean squared error for the covariate-specific cumulative-dynamic time-dependent ROC curve (over a sequence of covariate values) at the quartiles of the observed times ($t_{\text{Q1}}$, $t_{\text{Q2}}$ and $t_{\text{Q3}}$). Results are shown for the proposal presented in this paper with (w) and without (wo) model selection based on a double penalty approach.}
\label{ROC_RMSE_sim_w_wo_select}
\end{figure}

\begin{figure}[H]
\centering
\includegraphics[width=12cm]{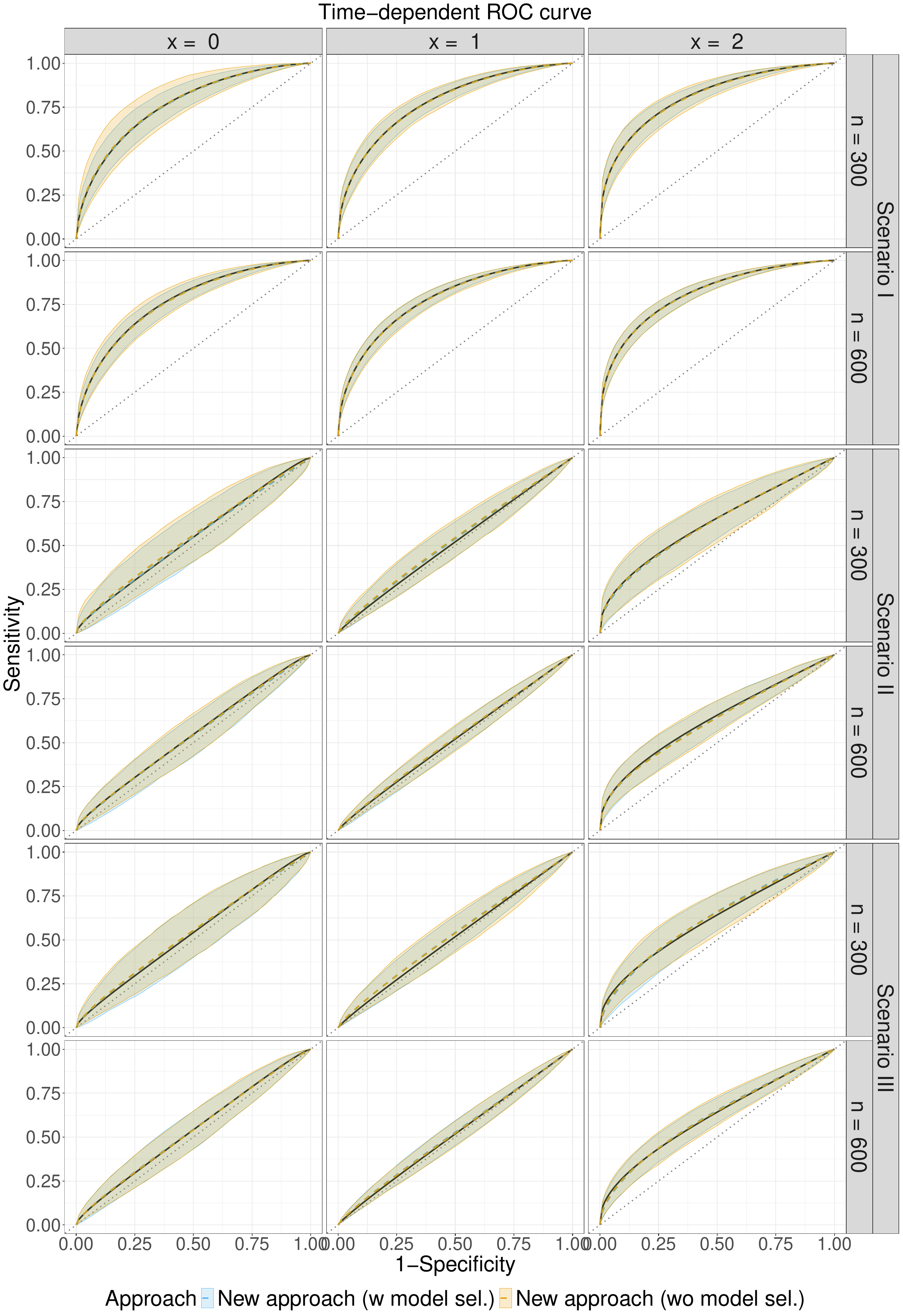}
\caption {True covariate-specific cumulative-dynamic time-dependent ROC curve (solid black line) versus the average of the estimated ROC curves (dotted coloured lines), evaluated at the median observed time ($t_{\text{Q2}}$) and for $x \in \{0, 1, 2\}$. The shaded areas are bands constructed using the pointwise 2.5\% and 97.5\% quantiles across simulations. Results are shown for the method proposed in this paper, with (w) and without (wo) model selection based on the double penalty approach.}
\label{ROC_Curves_sim_w_wo_select}
\end{figure}

\begin{figure}[H]
\centering
\includegraphics[width=12cm]{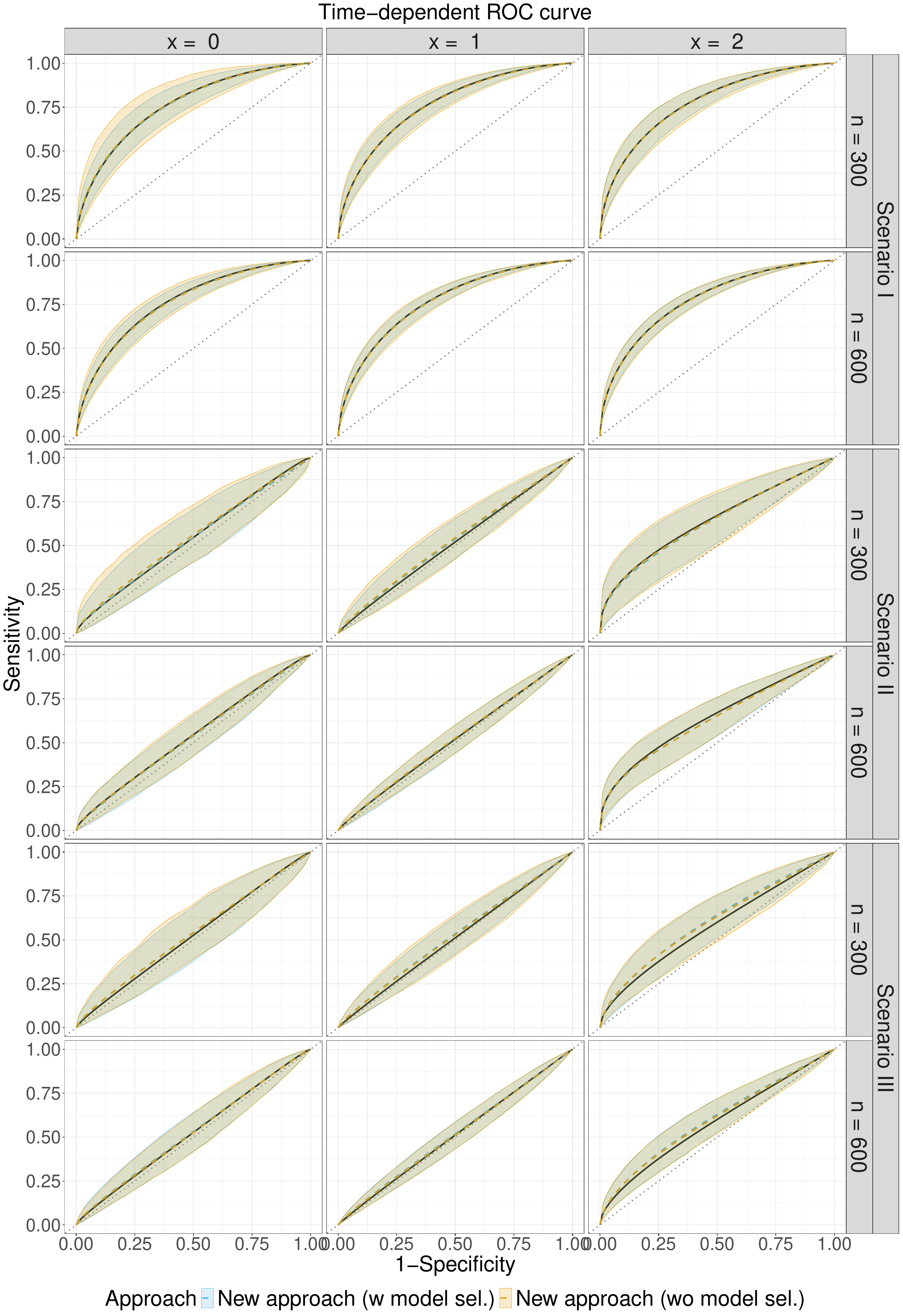}
\caption {True covariate-specific cumulative-dynamic time-dependent ROC curve (solid black line) versus the average of the estimated ROC curves (dotted coloured lines), evaluated at the first quartile of the observed time ($t_{\text{Q1}}$) and for $x \in \{0, 1, 2\}$. The shaded areas are bands constructed using the pointwise 2.5\% and 97.5\% quantiles across simulations. Results are shown for the method proposed in this paper, with (w) and without (wo) model selection based on the double penalty approach.}
\label{ROC_Curves_sim_w_wo_select_q1}
\end{figure}

\begin{figure}[H]
\centering
\includegraphics[width=12cm]{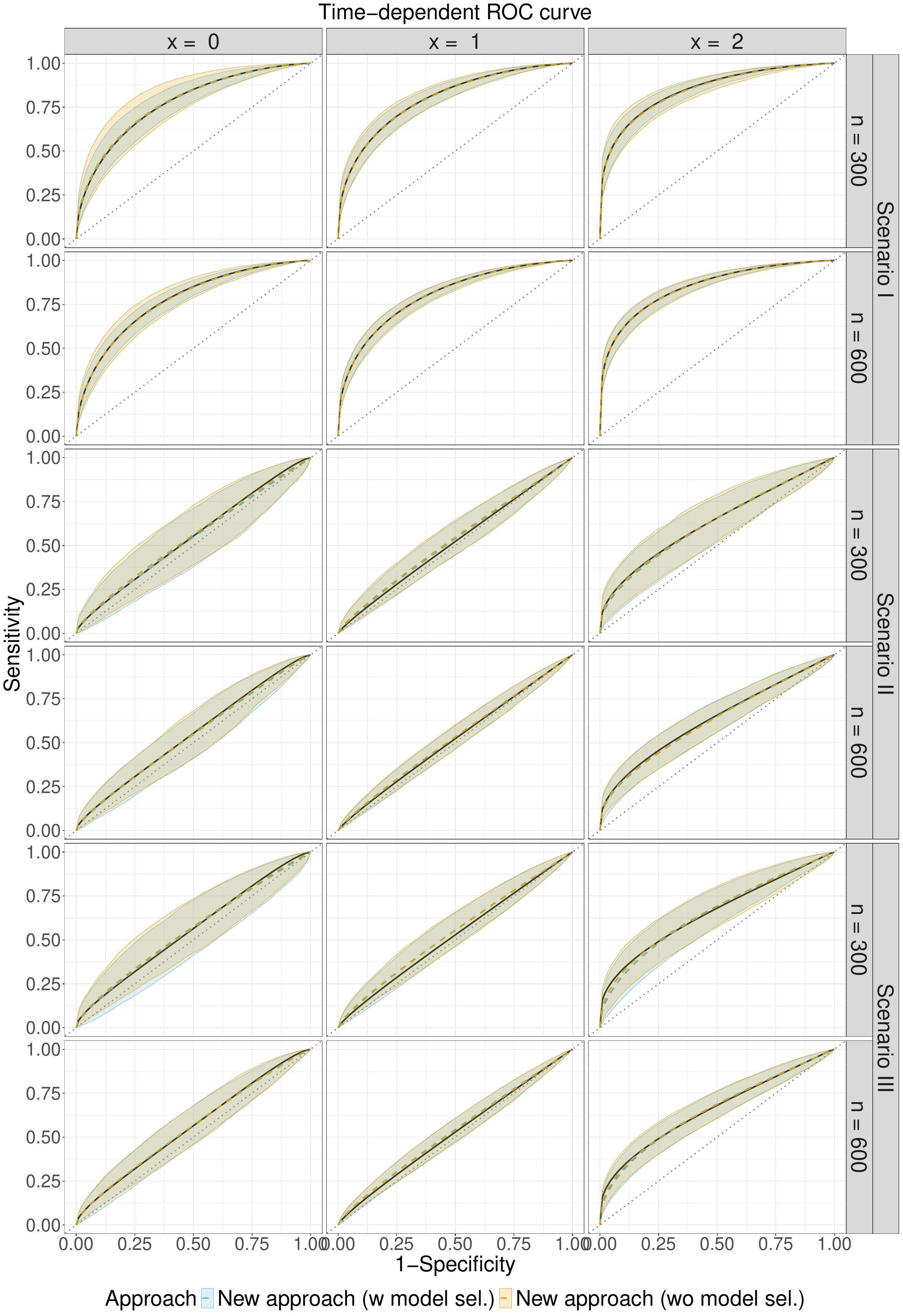}
\caption {True covariate-specific cumulative-dynamic time-dependent ROC curve (solid black line) versus the average of the estimated ROC curves (dotted coloured lines), evaluated at the third quartile of the observed time ($t_{\text{Q3}}$) and for $x \in \{0, 1, 2\}$. The shaded areas are bands constructed using the pointwise 2.5\% and 97.5\% quantiles across simulations. Results are shown for the method proposed in this paper, with (w) and without (wo) model selection based on the double penalty approach.}
\label{ROC_Curves_sim_w_wo_select_q3}
\end{figure}

\begin{figure}[H]
\centering
\includegraphics[width=12cm]{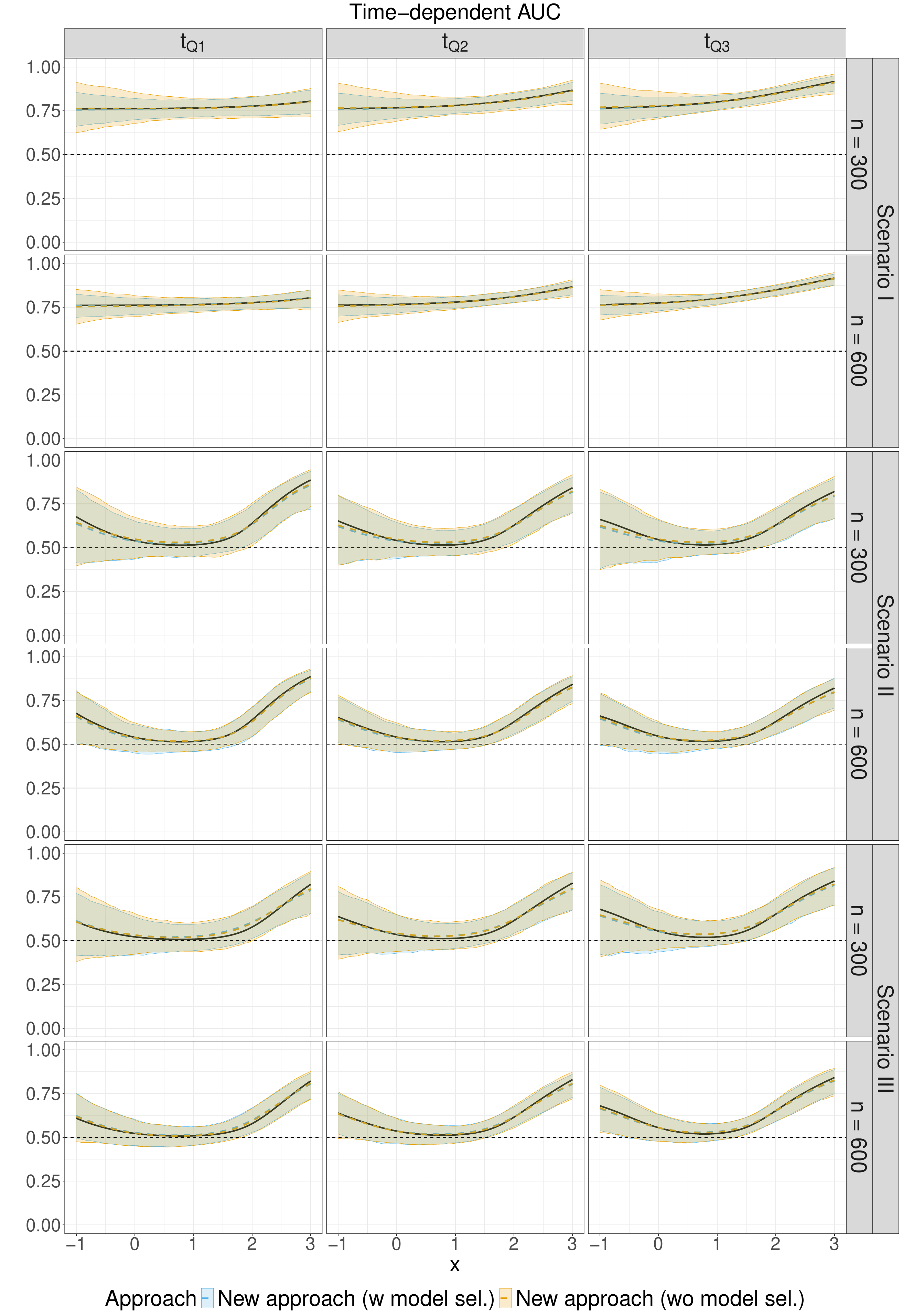}
\caption {True covariate-specific cumulative-dynamic time-dependent AUC (solid black line) versus the average of the estimated AUCs (dotted coloured lines), evaluated at the quartiles of the observed times ($t_{\text{Q1}}$, $t_{\text{Q2}}$, and $t_{\text{Q3}}$). The shaded areas are bands constructed using the pointwise 2.5\% and 97.5\% quantiles across simulations. Results are shown for the method proposed in this paper, with (w) and without (wo) model selection based on the double penalty approach.}
\label{AUC_sim_w_wo_select}
\end{figure}

\begin{figure}[H]
\centering
\includegraphics[width=12cm]{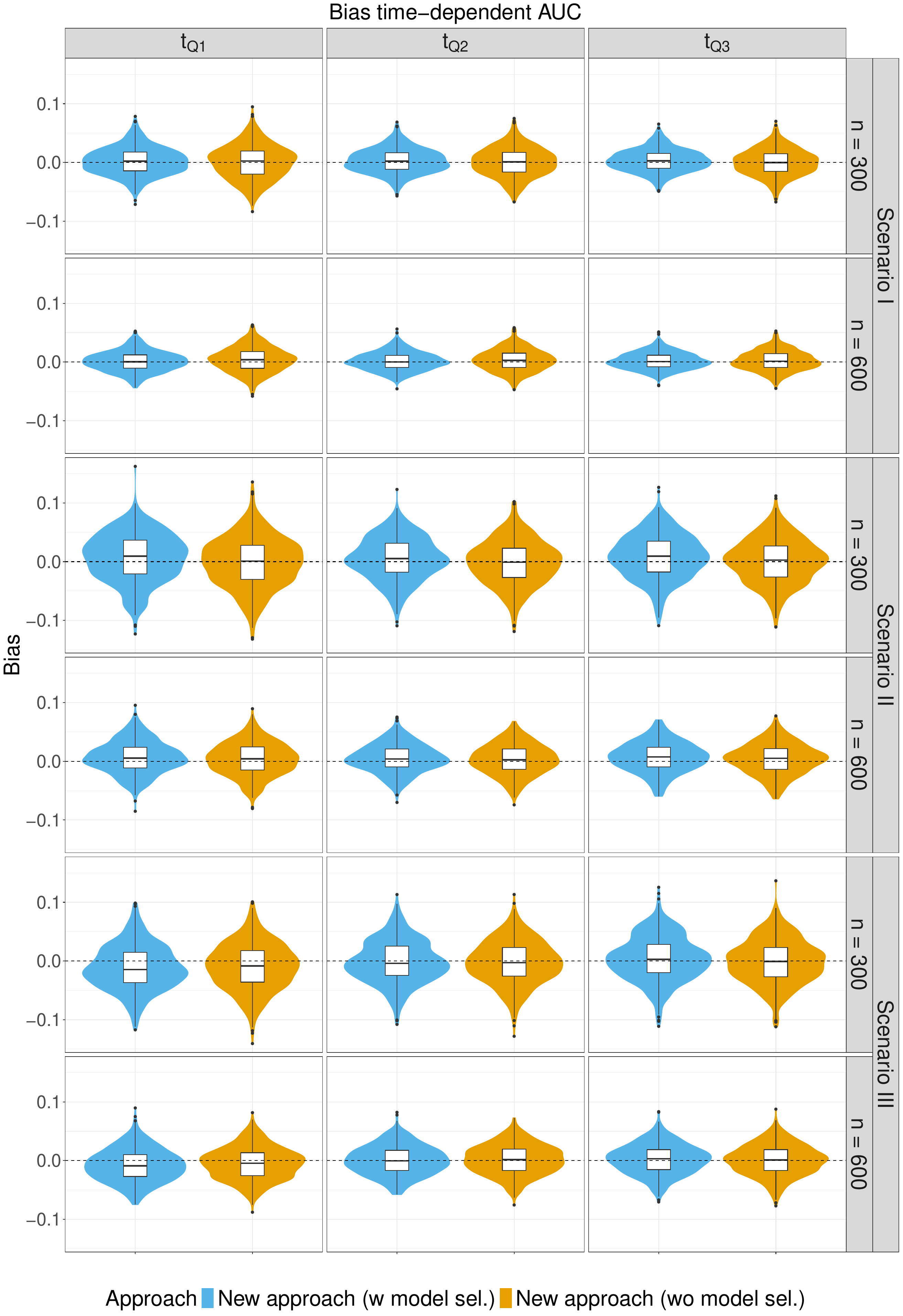}
\caption {Violin plot of the bias for the covariate-specific time-dependent cumulative AUC (over a sequence of covariate values) at the quartiles of the observed times ($t_{\text{Q1}}$, $t_{\text{Q2}}$, and $t_{\text{Q3}}$). Results are shown for the method proposed in this paper, with (w) and without (wo) model selection based on the double penalty approach.}
\label{Bias_AUC_sim_w_wo_select}
\end{figure}

\begin{figure}[H]
\centering
\includegraphics[width=12cm]{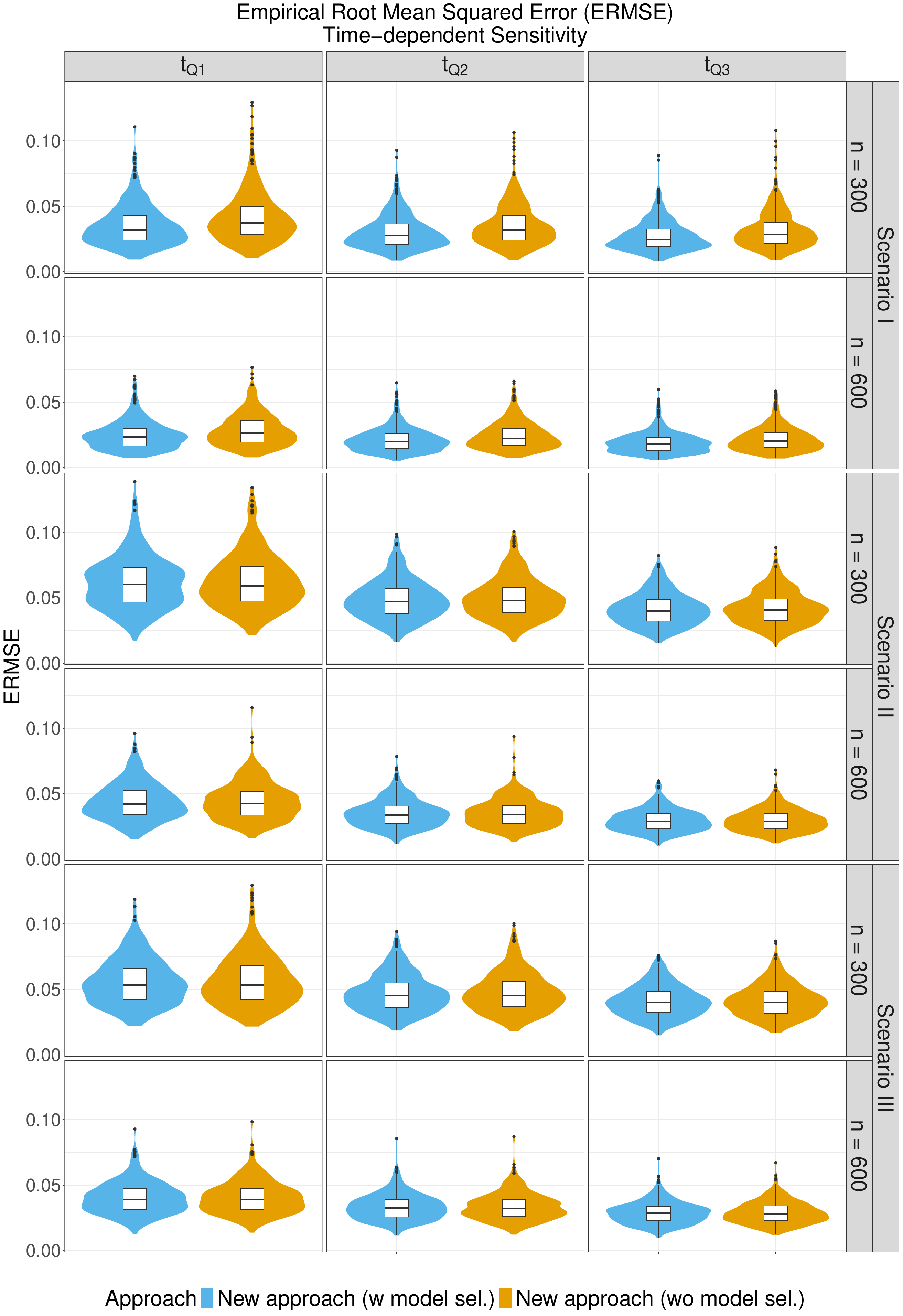}
\caption {Violin plot of the empirical root mean squared error for the covariate-specific time-dependent cumulative \textit{Sensitivity} (over a sequence of covariate values) at the quartiles of the observed times ($t_{\text{Q1}}$, $t_{\text{Q2}}$, and $t_{\text{Q3}}$). Results are shown for the method proposed in this paper, with (w) and without (wo) model selection based on the double penalty approach.}
\label{TPR_RMSE_sim_w_wo_select}
\end{figure}

\begin{figure}[H]
\centering
\includegraphics[width=12cm]{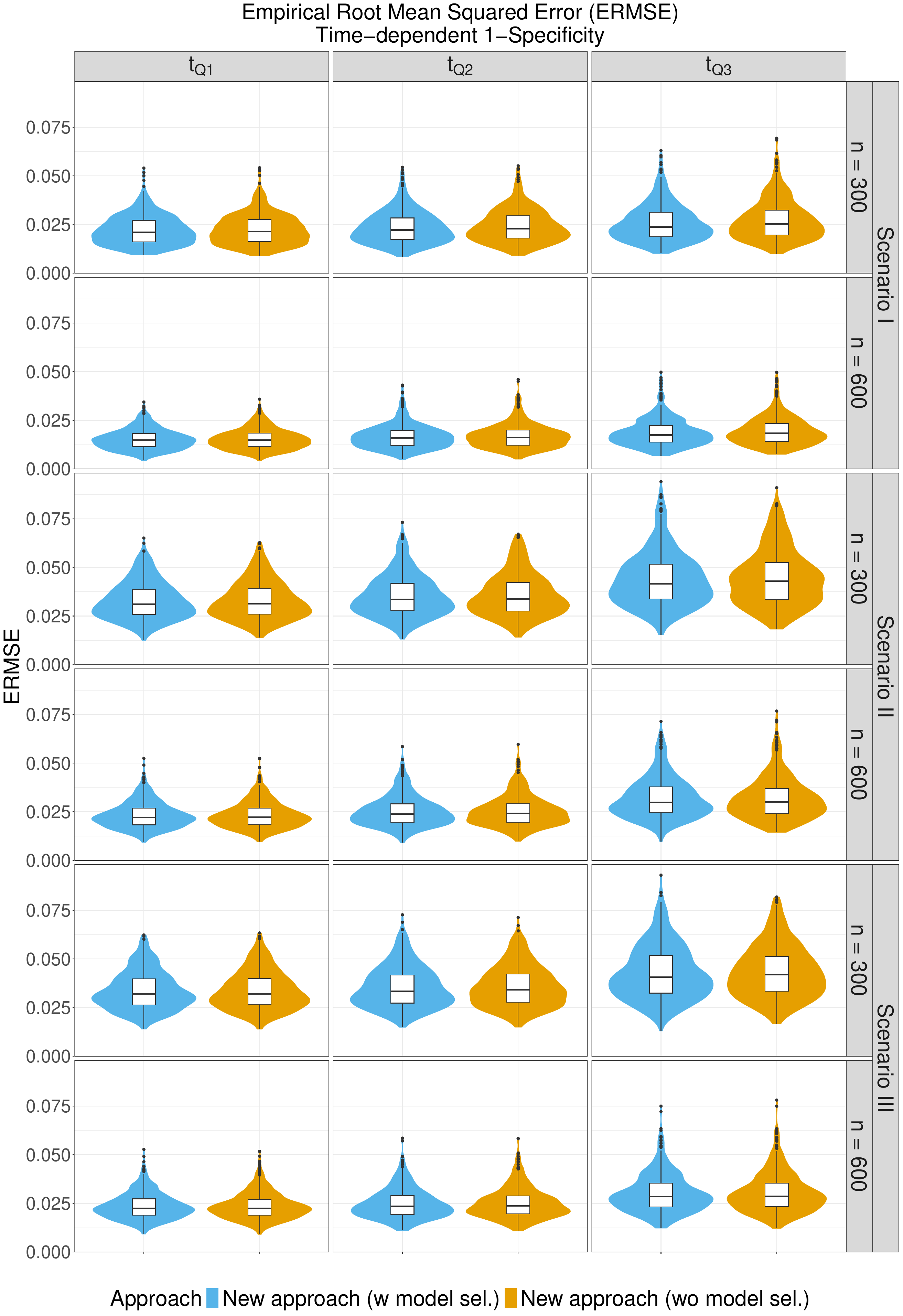}
\caption {Violin plot of the empirical root mean squared error for the covariate-specific time-dependent dynamic 1-\textit{Specificity} (over a sequence of covariate values) at the quartiles of the observed times ($t_{\text{Q1}}$, $t_{\text{Q2}}$, and $t_{\text{Q3}}$). Results are shown for the method proposed in this paper, with (w) and without (wo) model selection based on the double penalty approach.}
\label{FPR_RMSE_sim_w_wo_select}
\end{figure}

\subsection{New Approach: Break Points for the Piece-wise Exponential Model}\label{supp:simulation:diff_timepoints}
\begin{figure}[H]
\centering
\includegraphics[width = 12cm]{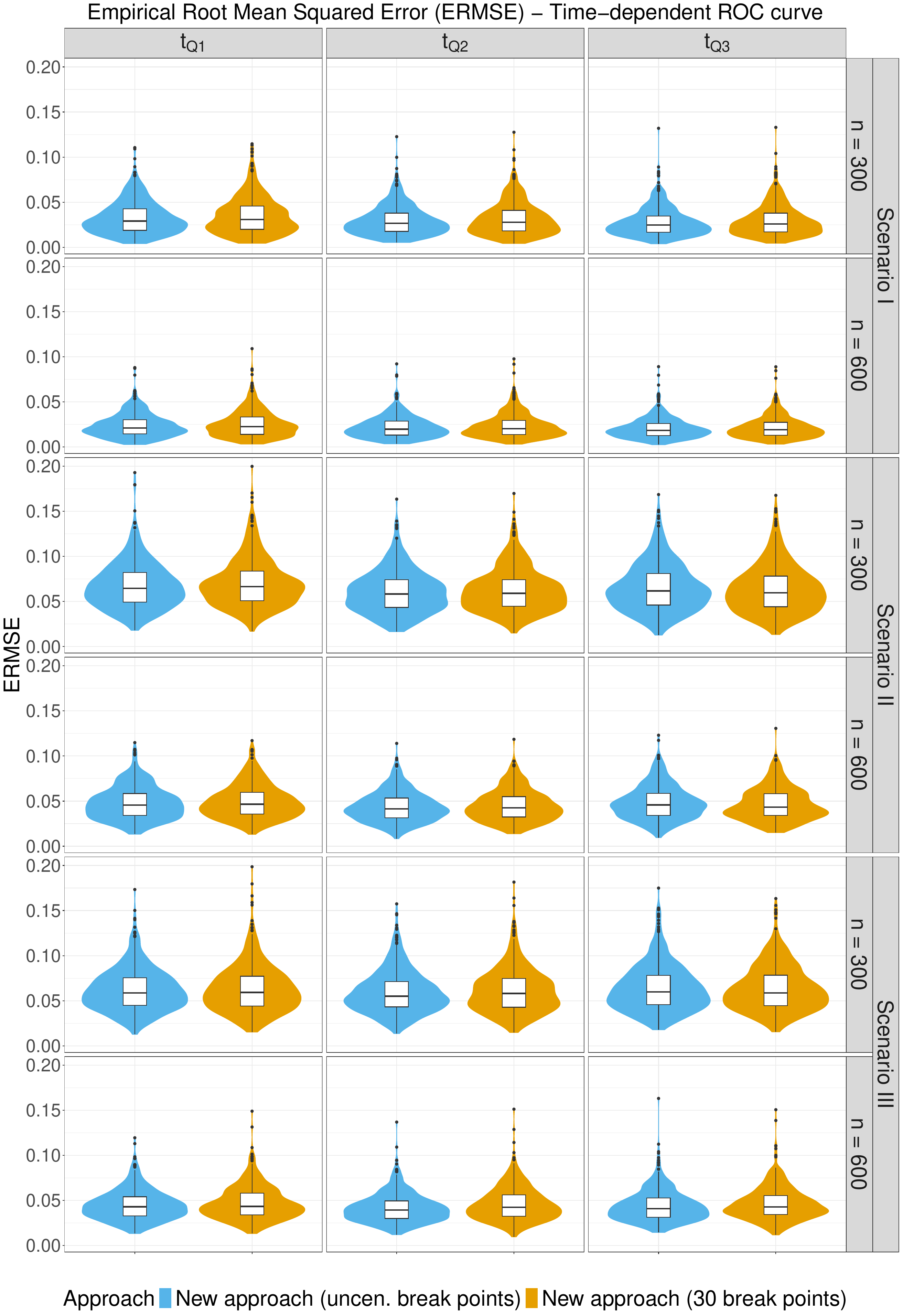}
\caption {Violin plot of the empirical root mean squared error for the covariate-specific cumulative-dynamic time-dependent ROC curve (over a sequence of covariate values), evaluated at the quartiles of the observed times ($t_{\text{Q1}}$, $t_{\text{Q2}}$, and $t_{\text{Q3}}$).  Results are shown for the method proposed in this paper, using break points for the follow-up time in the piece-wise exponential approach set either at the observed uncensored times or at 30 equally spaced values.}
\label{ROC_RMSE_sim_timepoints}
\end{figure}

\begin{figure}[H]
\centering
\includegraphics[width=12cm]{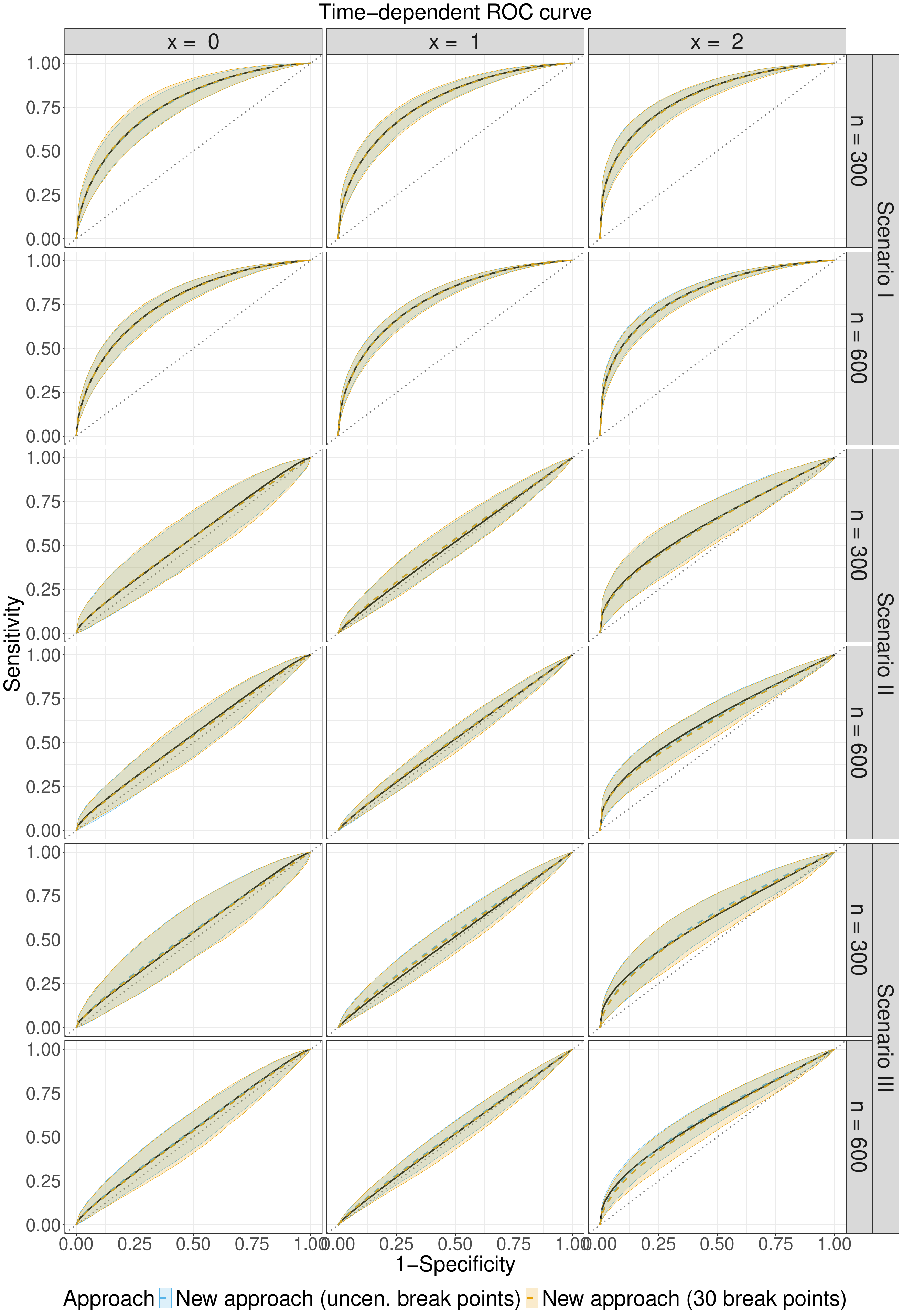}
\caption {True covariate-specific cumulative-dynamic time-dependent ROC curve (solid black line) versus the average of the estimated ROC curves (dotted coloured lines), evaluated at the median observed time ($t_{\text{Q2}}$) and for $x \in \{0, 1, 2\}$. The shaded areas are bands constructed using the pointwise 2.5\% and 97.5\% quantiles across simulations.  Results are shown for the method proposed in this paper, using break points for the follow-up time in the piece-wise exponential approach set either at the observed uncensored times or at 30 equally spaced values.}
\label{ROC_Curves_sim_timepoints}
\end{figure}

\begin{figure}[H]
\centering
\includegraphics[width=12cm]{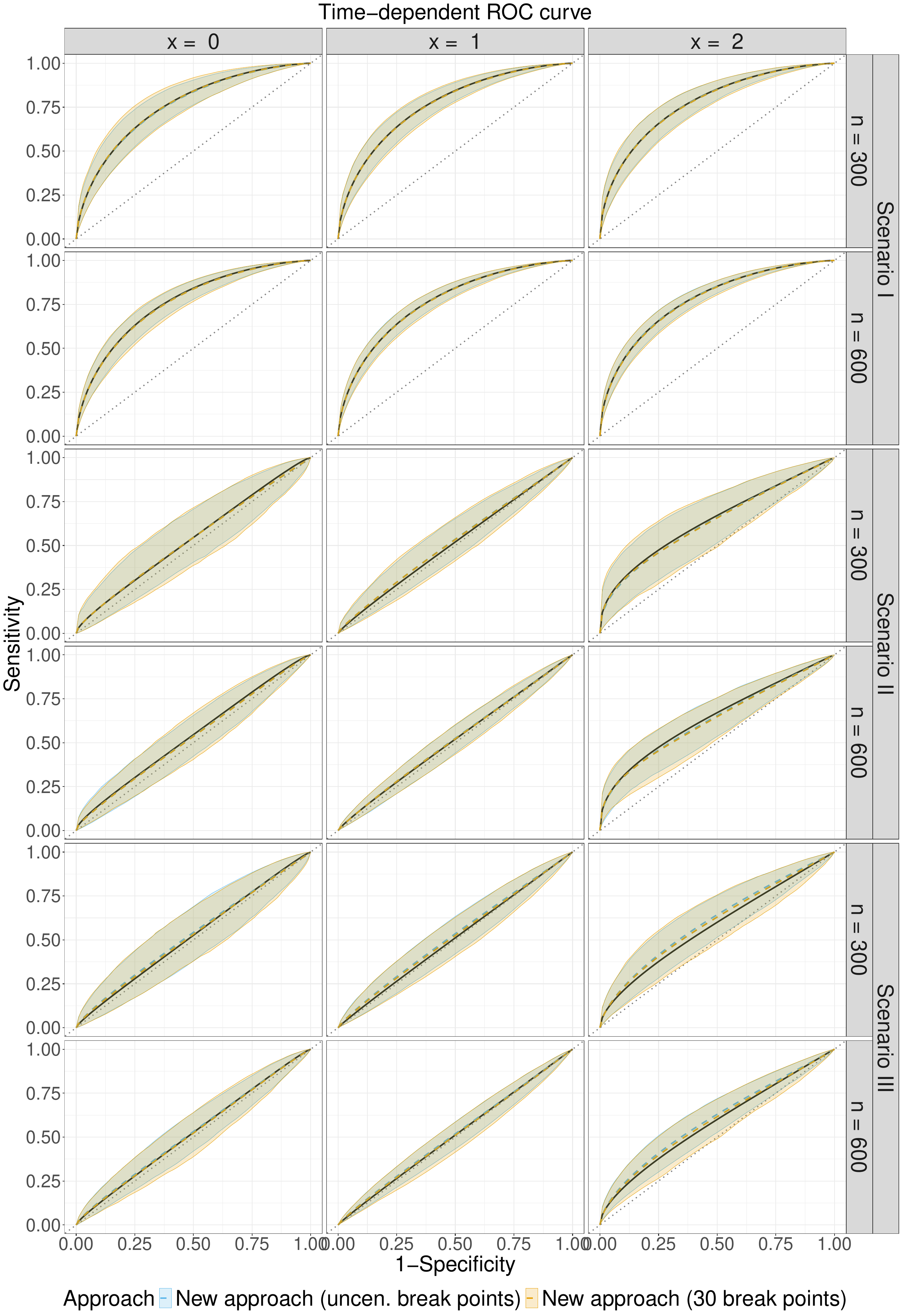}
\caption {True covariate-specific cumulative-dynamic time-dependent ROC curve (solid black line) versus the average of the estimated ROC curves (dotted coloured lines), evaluated at the first quartile of the observed time ($t_{\text{Q1}}$) and for $x \in \{0, 1, 2\}$. The shaded areas are bands constructed using the pointwise 2.5\% and 97.5\% quantiles across simulations. Results are shown for the method proposed in this paper, using break points for the follow-up time in the piece-wise exponential approach set either at the observed uncensored times or at 30 equally spaced values.}
\label{ROC_Curves_sim_timepoints_q1}
\end{figure}

\begin{figure}[H]
\centering
\includegraphics[width=12cm]{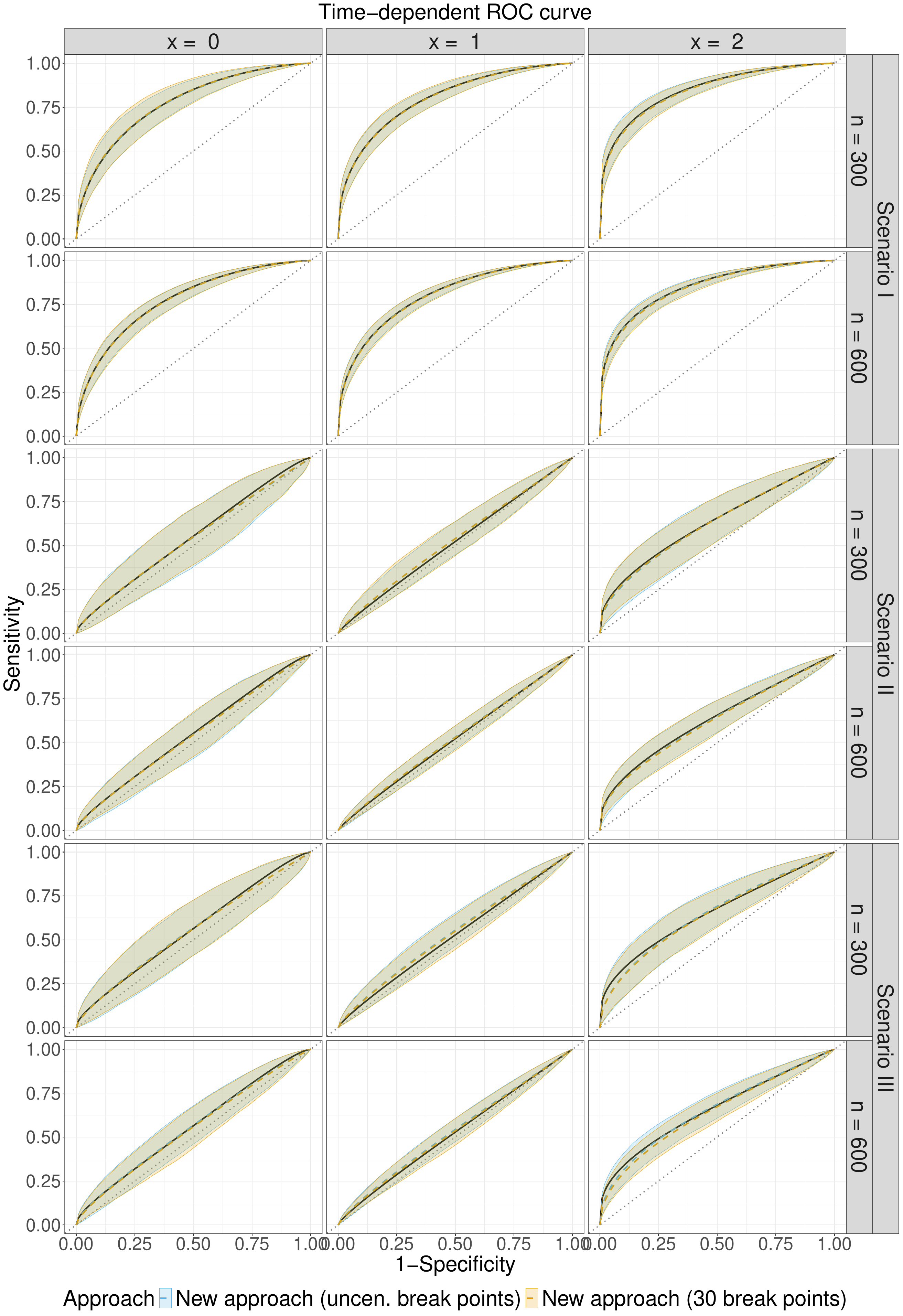}
\caption {True covariate-specific cumulative-dynamic time-dependent ROC curve (solid black line) versus the average of the estimated ROC curves (dotted coloured lines), evaluated at the third quartile of the observed time ($t_{\text{Q3}}$) and for $x \in \{0, 1, 2\}$. The shaded areas are bands constructed using the pointwise 2.5\% and 97.5\% quantiles across simulations. Results are shown for the method proposed in this paper, using break points for the follow-up time in the piece-wise exponential approach set either at the observed uncensored times or at 30 equally spaced values.}
\label{ROC_Curves_sim_timepoints_q3}
\end{figure}

\begin{figure}[H]
\centering
\includegraphics[width=12cm]{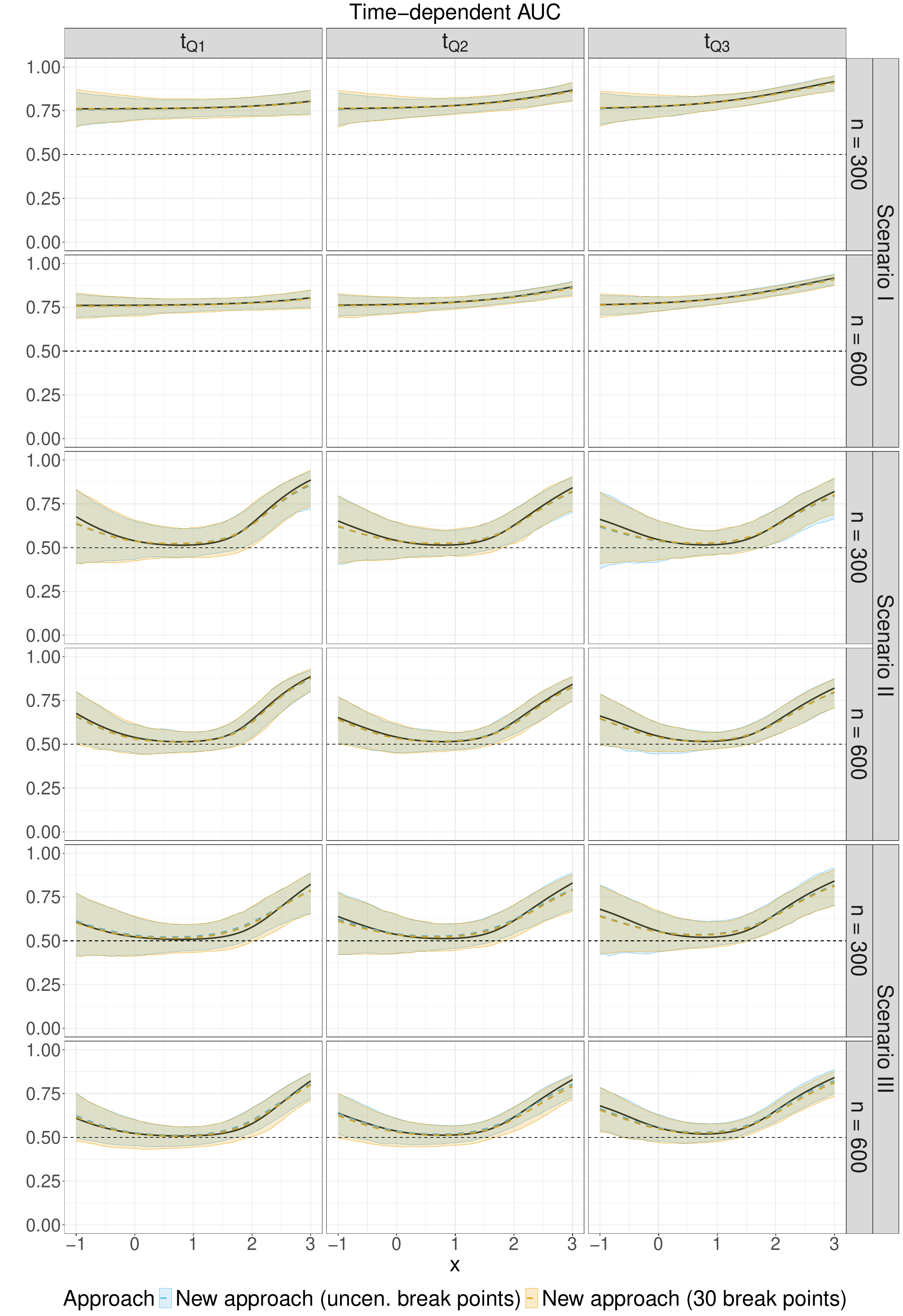}
\caption {True covariate-specific cumulative-dynamic time-dependent AUC (solid black line) versus the average of the estimated AUCs (dotted coloured lines) at the quartiles of the observed times ($t_{\text{Q1}}$, $t_{\text{Q2}}$, and $t_{\text{Q3}}$). The shaded areas are bands constructed using the pointwise 2.5\% and 97.5\% quantiles across simulations. Results are shown for the method proposed in this paper, using break points for the follow-up time in the piece-wise exponential approach set either at the observed uncensored times or at 30 equally spaced values.}
\label{AUC_sim_timepoints}
\end{figure}

\begin{figure}[H]
\centering
\includegraphics[width=12cm]{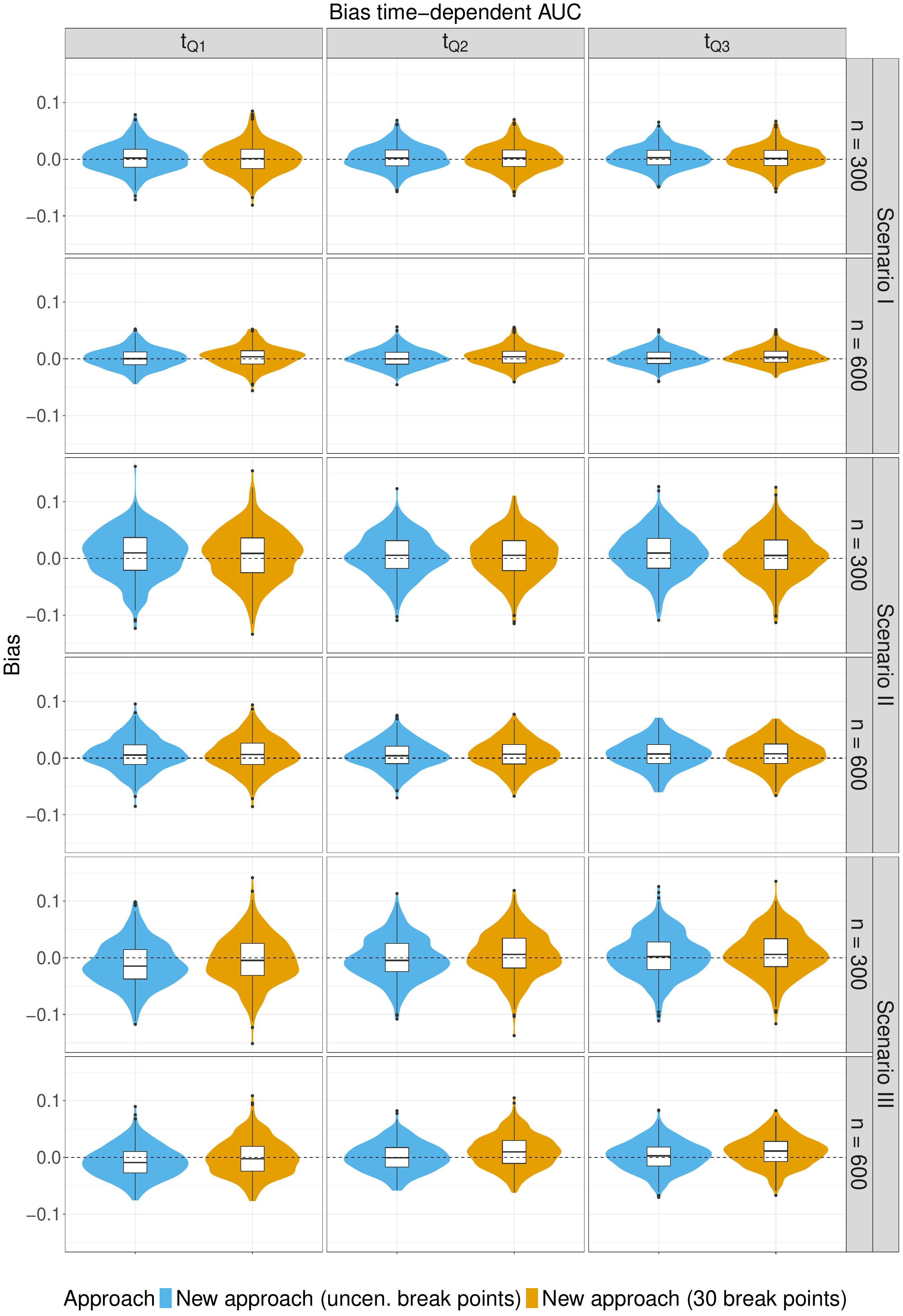}
\caption {Violin plot of the bias for the covariate-specific time-dependent cumulative AUC (over a sequence of covariate values), evaluated at the quartiles of the observed times ($t_{\text{Q1}}$, $t_{\text{Q2}}$, and $t_{\text{Q3}}$). Results are shown for the method proposed in this paper, using break points for the follow-up time in the piece-wise exponential approach set either at the observed uncensored times or at 30 equally spaced values.}
\label{Bias_AUC_sim_timepoints}
\end{figure}

\begin{figure}[H]
\centering
\includegraphics[width=12cm]{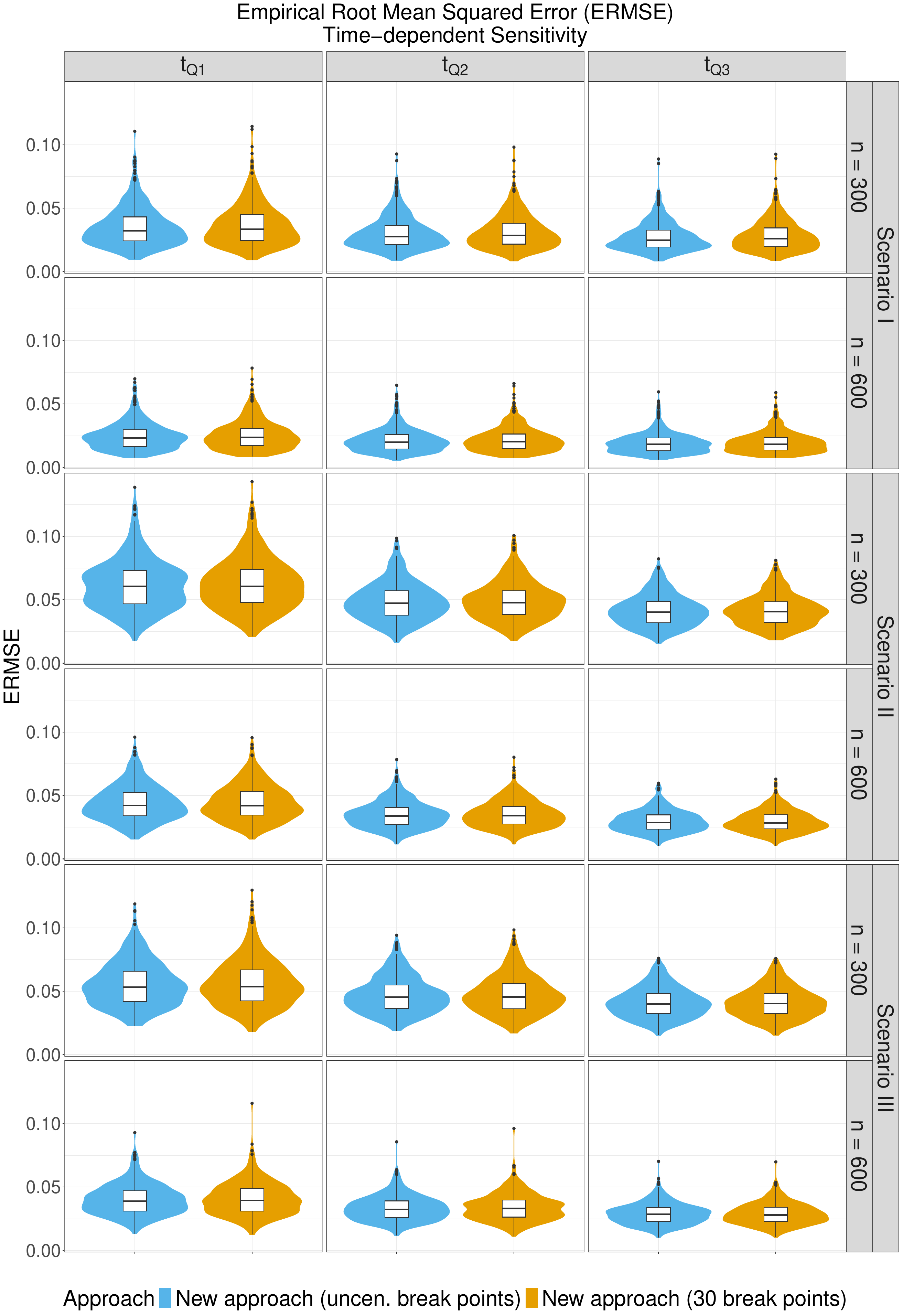}
\caption {Violin plot of the empirical root mean squared error for the covariate-specific time-dependent cumulative \textit{Sensitivity} (over a sequence of covariate values), evaluated at the quartiles of the observed times ($t_{\text{Q1}}$, $t_{\text{Q2}}$, and $t_{\text{Q3}}$).  Results are shown for the method proposed in this paper, using break points for the follow-up time in the piece-wise exponential approach set either at the observed uncensored times or at 30 equally spaced values.}
\label{TPR_RMSE_sim_timepoints}
\end{figure}

\begin{figure}[H]
\centering
\includegraphics[width=12cm]{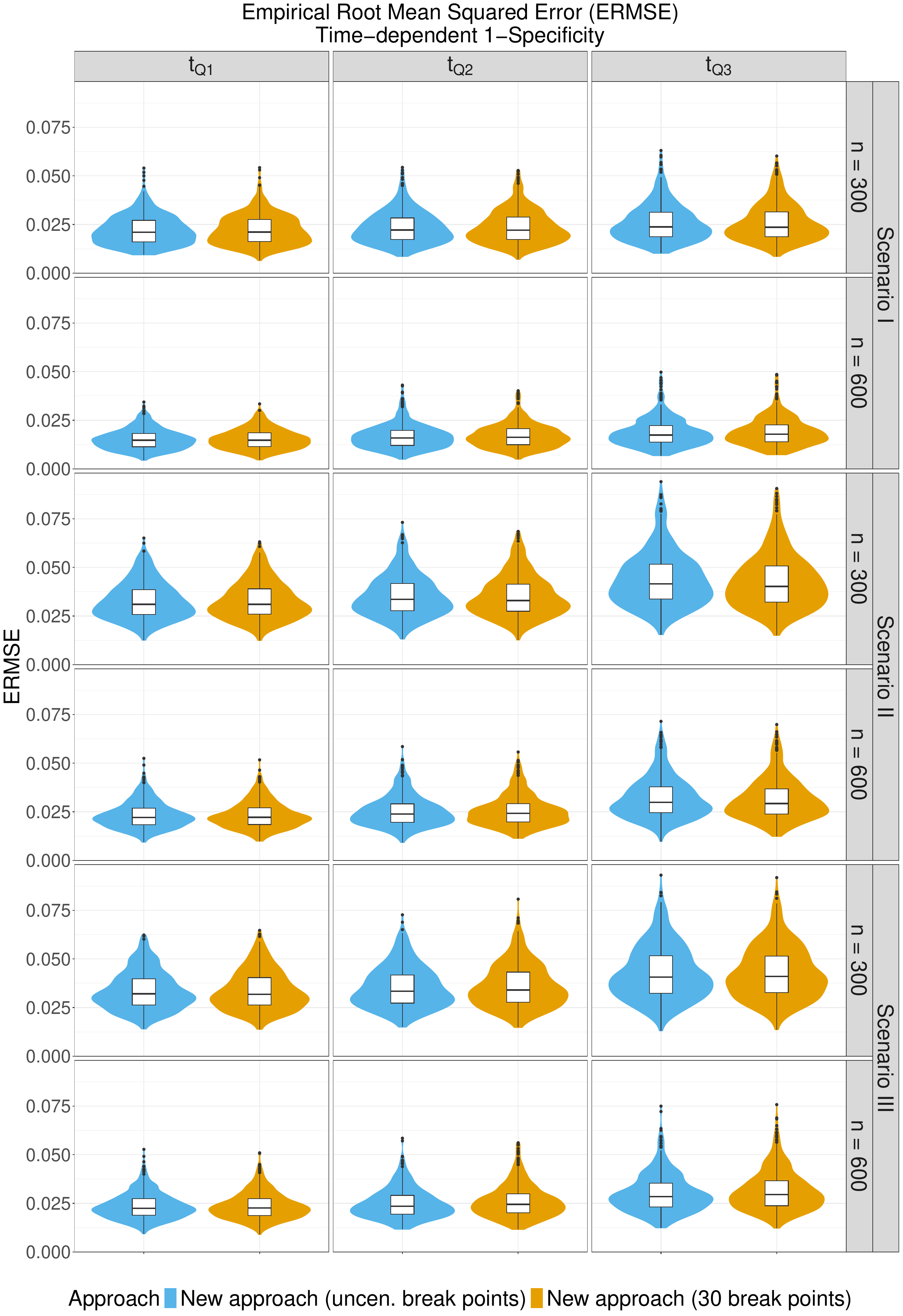}
\caption {Violin plot of the empirical root mean squared error for the covariate-specific time-dependent dynamic 1-\textit{Specificity} (over a sequence of covariate values), evaluated at the quartiles of the observed times ($t_{\text{Q1}}$, $t_{\text{Q2}}$, and $t_{\text{Q3}}$). Results are shown for the method proposed in this paper, using break points for the follow-up time in the piece-wise exponential approach set either at the observed uncensored times or at 30 equally spaced values.}
\label{FPR_RMSE_sim_timepoints}
\end{figure}

\section{Real Data Application: Extra Results}\label{supp:app}

\begin{figure}[H]
\centering
\includegraphics[width=15cm]{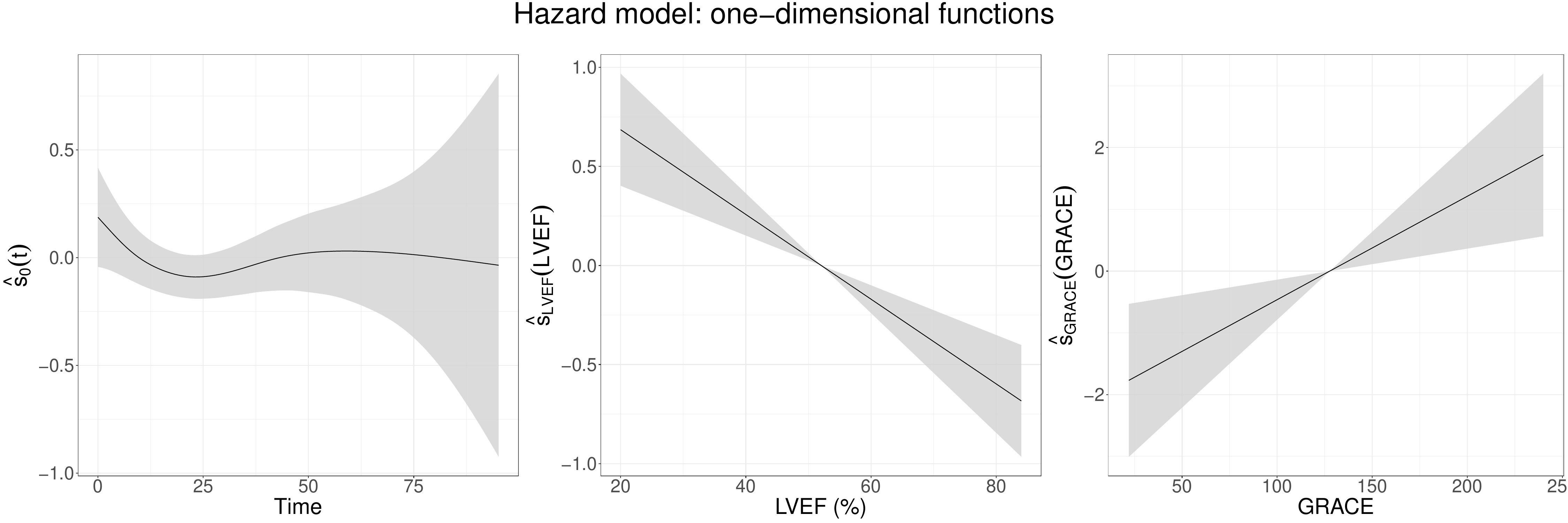}
\vspace{5mm}

\includegraphics[width=15cm]{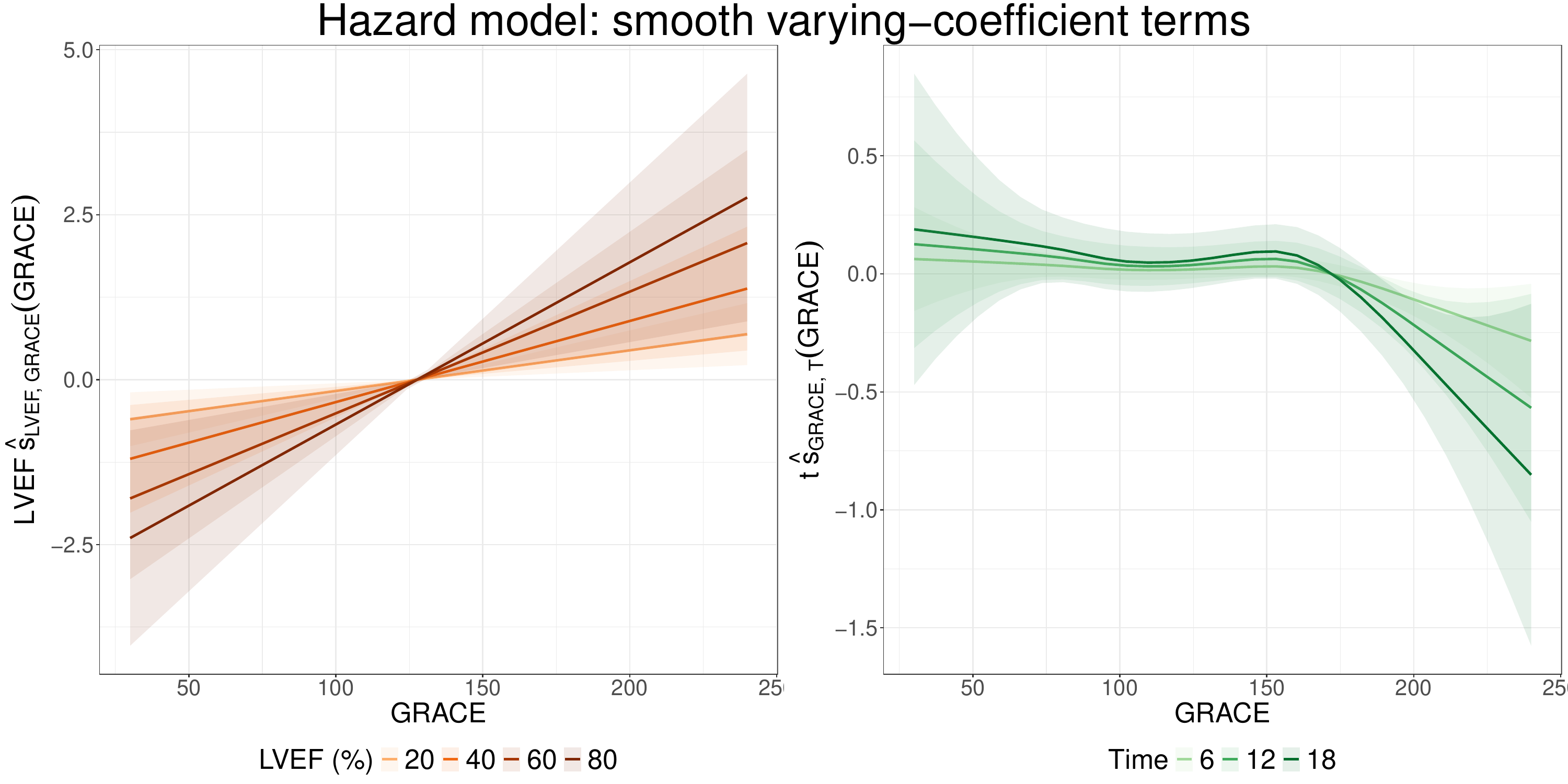}
\caption {For the real data application and the model for the conditional hazard function: Top row: Estimated one-dimensional smooth functions (black lines). Grey shaded areas represent pointwise 95\% confidence intervals. Bottom row: Estimated smooth varying-coefficient terms of the GRACE risk score for different values of the LVEF (\%) (left panel) and times (months post-discharge; right panel). Coloured areas represent pointwise 95\% confidence intervals.}
\label{hazard_model_effects}
\end{figure}
\begin{figure}[H]
\centering
\includegraphics[width=14cm]{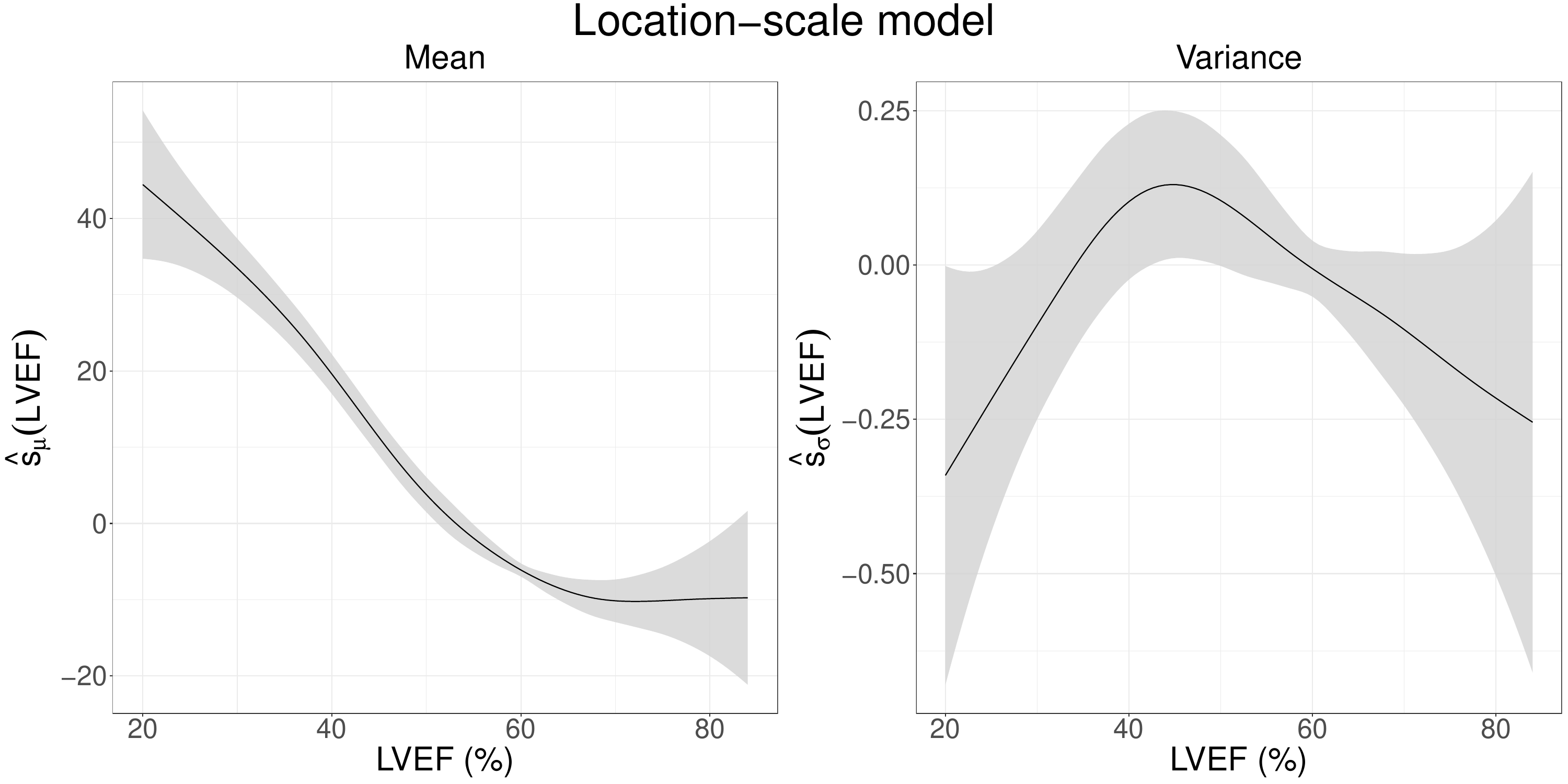}
\caption{For the real data application and the location-scale regression model for the GRACE score: Estimated one-dimensional smooth functions (black lines) for the LVEF (\%). Grey shaded areas correspond to pointwise 95\% confidence intervals.}
\label{loc_scale_model_effects}
\end{figure}
\begin{figure}
\centering
\includegraphics[width=13.5cm]{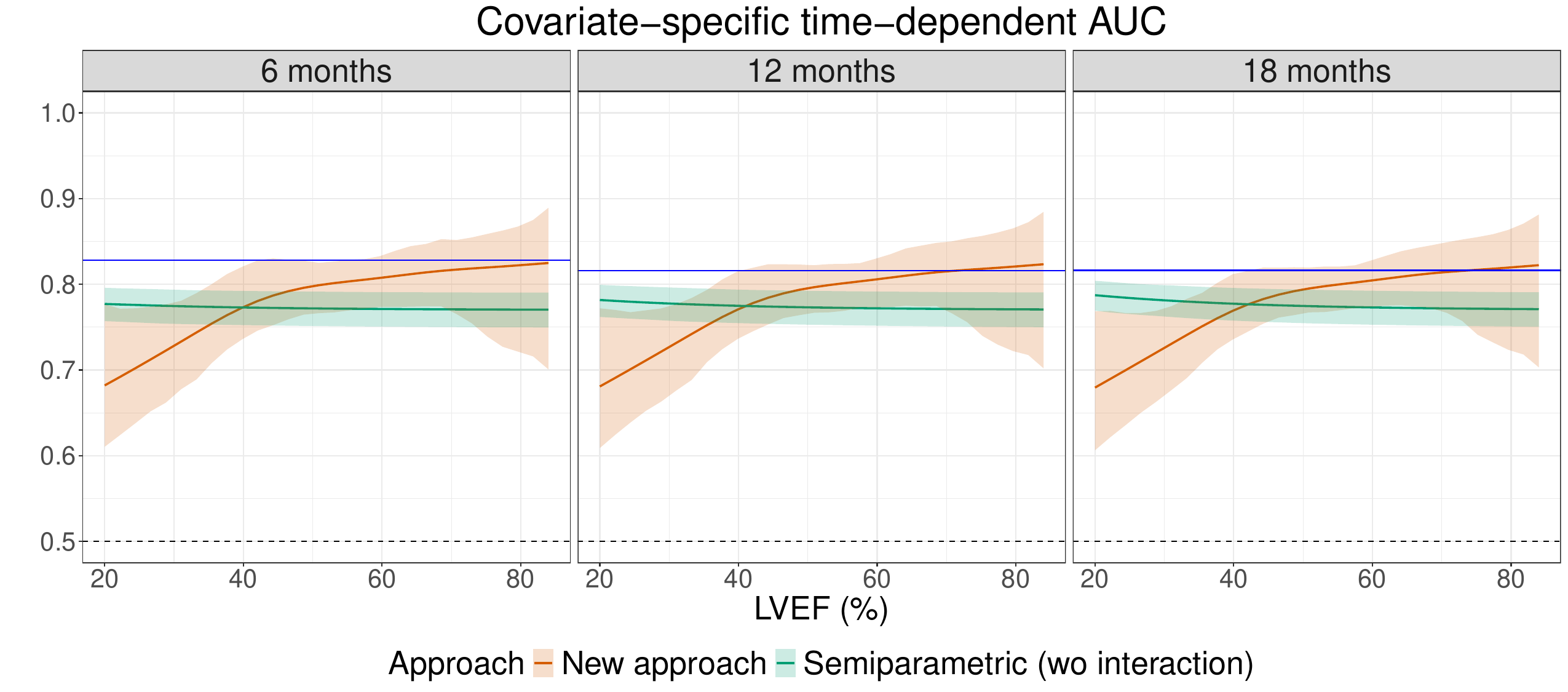}
\includegraphics[width=13.5cm]{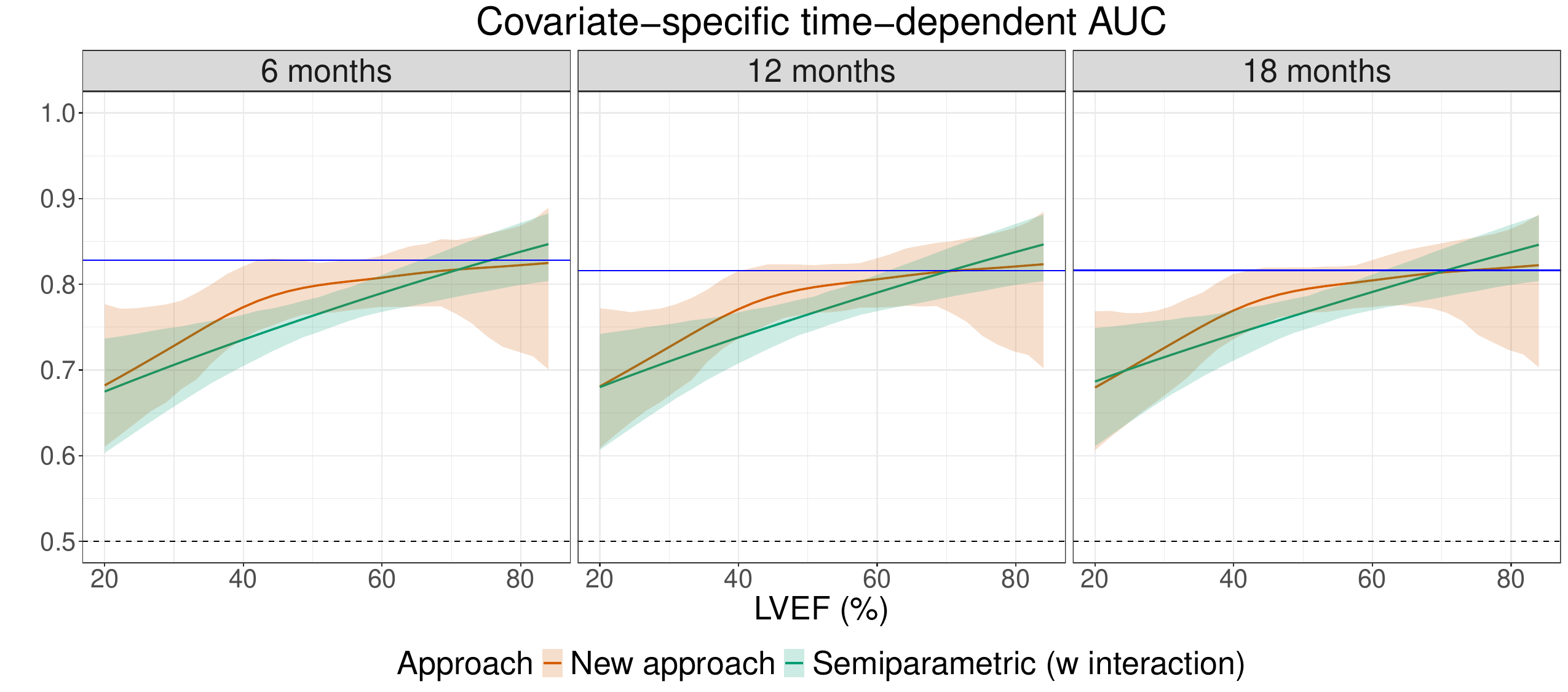}
\includegraphics[width=13.5cm]{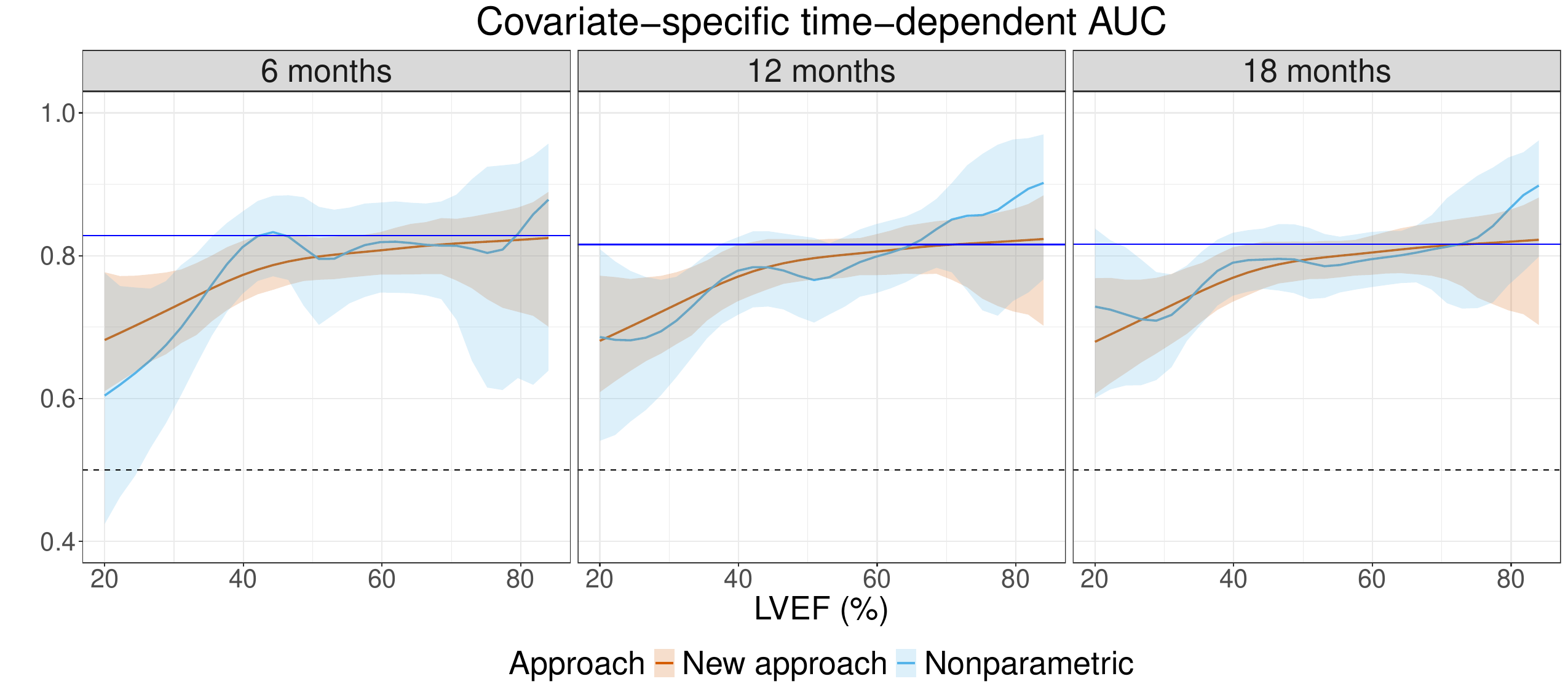}
\caption {Estimated covariate-specific cumulative-dynamic time-dependent AUC of the GRACE score adjusted for LVEF(\%) with 95\% pointwise bootstrap confidence bands at $t = 6, 12$, and $18$ months post-discharge. `New approach' corresponds to the method proposed in this paper,  `Semiparametric' refers to the approach by~\cite{xiao08}, and `Nonparametric' to the smoothed nonparametric method by~\cite{MX16}. For the `Semiparametric' method, the top and middle rows display results without and with the linear interaction between LVEF and the GRACE score in the Cox model's linear predictor, respectively. The blue horizontal lines correspond to the estimated pooled cumulative-dynamic time-dependent AUCs using the method by \cite{Uno2007}.}
\label{AUC_Real_Data_all_supp}
\end{figure}

\clearpage
\putbib[ctimeROC]
\end{bibunit}
\end{document}